\documentclass[10pt,preprint]{emulateapj} 
\begin{document}
\title{The ISLAndS project II: The Lifetime Star Formation Histories of Six Andromeda dSphs\altaffilmark{1}}

\author{
Evan D. Skillman\altaffilmark{2},
Matteo Monelli\altaffilmark{3,4},
Daniel R. Weisz\altaffilmark{5,6},
Sebastian L. Hidalgo\altaffilmark{3,4},
Antonio Aparicio\altaffilmark{4,3},
Edouard J. Bernard\altaffilmark{7},
Michael Boylan-Kolchin\altaffilmark{8}, 
Santi Cassisi\altaffilmark{9},
Andrew A. Cole\altaffilmark{10},
Andrew E. Dolphin\altaffilmark{11},
Henry C. Ferguson\altaffilmark{12},
Carme Gallart\altaffilmark{3,4},
Mike J. Irwin\altaffilmark{13},
Nicolas F. Martin\altaffilmark{14,15},
Clara E. Mart{\'i}nez-V{\'a}zquez\altaffilmark{3,4},
Lucio Mayer\altaffilmark{16,17},
Alan W. McConnachie\altaffilmark{18},
Kristen B. W. McQuinn\altaffilmark{8},
Julio F. Navarro\altaffilmark{19}, and
Peter B. Stetson\altaffilmark{18}}

\altaffiltext{1}{Based on observations made with the NASA/ESA Hubble Space Telescope, 
obtained at the Space Telescope Science Institute, which is operated by the 
Association of Universities for Research in Astronomy, Inc., under NASA contract NAS 
5-26555. These observations are associated with programs \#13028, 13739.}
\altaffiltext{2}{Minnesota Institute for Astrophysics, University of Minnesota, 
Minneapolis, MN, USA; skillman@astro.umn.edu}
\altaffiltext{3}{Instituto de Astrof\'\i sica de Canarias (IAC). V\'\i a L\'actea s/n.
E38205 - La Laguna, Tenerife, Canary Islands, Spain;
monelli@iac.es, shidalgo@iac.es, aaj@iac.es, carme@iac.es, clara.marvaz@gmail.com}
\altaffiltext{4}{Department of Astrophysics, University of La Laguna. V\'\i a L\'actea s/n.
E38206 - La Laguna, Tenerife, Canary Islands, Spain}
\altaffiltext{5}{Astronomy Department, Box 351580, University of Washington, Seattle, WA, USA;
dweisz@uw.edu}
%\altaffiltext{6}{Department of Astronomy, University of California at Santa Cruz,
%1156 High Street, Santa Cruz, CA, 95064}
\altaffiltext{6}{Hubble Fellow}
\altaffiltext{7}{Laboratoire Lagrange (UMR7293), Observatoire de la C\^ote d’Azur, 
F-06304 Nice, France; ebernard@oca.eu}
%Institute for Astronomy, University of Edinburgh, Royal
%Observatory, Blackford Hill, Edinburgh EH9 3HJ, UK; ejb@roe.ac.uk}
\altaffiltext{8}{Astronomy Department, University of Texas, Austin, TX, USA;
mbk@astro.as.utexas.edu, kmcquinn@astro.as.utexas.edu}
\altaffiltext{9}{INAF-Osservatorio Astronomico di Collurania,
Teramo, Italy; cassisi@oa-teramo.inaf.it}
\altaffiltext{10}{School of Physical Sciences, University of Tasmania,
Hobart, Tasmania, Australia; andrew.cole@utas.edu.au}
\altaffiltext{11}{Raytheon; 1151 E. Hermans Rd., Tucson, AZ 85706, USA;
adolphin@raytheon.com}
\altaffiltext{12}{Space Telescope Science Institute, 3700 San Martin Drive, Baltimore, MD 21218, USA; 
ferguson@stsci.edu}
\altaffiltext{13}{University of Cambridge, Madingley Road, Cambridge CB3 0HA, UK; mike@ast.cam.ac.uk}
\altaffiltext{14}{Observatoire astronomique de Strasbourg, Université de Strasbourg, CNRS, UMR 7550, 11 rue de l'Université, 
F-67000 Strasbourg, France; nicolas.martin@astro.unistra.fr} 
\altaffiltext{15}{Max-Planck-Institut für Astronomie, Königstuhl 17, D-69117 Heidelberg, Germany}
\altaffiltext{16}{Institut f\"ur Theoretische Physik, University of Zurich,
Z\"urich, Switzerland; lucio@physik.unizh.ch}
\altaffiltext{17}{Department of Physics, Institut f\"ur Astronomie,
ETH Z\"urich, Z\"urich, Switzerland; lucio@phys.ethz.ch}
\altaffiltext{18}{Dominion Astrophysical Observatory, Herzberg Institute of
Astrophysics, National Research Council, 5071 West Saanich Road, Victoria,
British Columbia V9E 2E7, Canada; peter.stetson@nrc-cnrc.gc.ca}
\altaffiltext{19} {Department of Physics and Astronomy, University of Victoria, BC V8P 5C2, Canada;
jfn@uvic.ca}
%\altaffiltext{18}{Kapteyn Astronomical Institute, University of Groningen,
%    Groningen, Netherlands; etolstoy@astro.rug.nl}

\begin{abstract}

The Initial Star formation and Lifetimes of Andromeda Satellites (ISLAndS) project
uses {\it Hubble Space Telescope} imaging to study a 
representative sample of six Andromeda dSph satellite companion galaxies. 
The main goal of the program is to determine whether the star formation histories (SFHs) of
the Andromeda dSph satellites demonstrate significant statistical differences from those of the Milky Way,
which may be attributable to the different properties of their local environments. 
Our observations reach the oldest main sequence turn-offs, allowing a time 
resolution at the oldest ages of $\sim 1$ Gyr, which is comparable to the best achievable 
resolution in the MW satellites.  We find that the six dSphs present a
variety of SFHs that are not strictly correlated with luminosity or present distance
from M31. Specifically, we find a significant range in quenching times ($\tau_{q}$, lookback times from 
9 to 6 Gyr), but with all quenching
times more than $\sim$6 Gyr ago.  In agreement with
observations of Milky Way companions of similar mass, there is no evidence of complete 
quenching of star formation by the cosmic UV background responsible for reionization, but
the possibility of a degree of quenching at reionization cannot be ruled out.  We do not find 
significant differences between the SFHs of the three members of the vast, thin plane of
satellites and the three off-plane dSphs.  
The primary difference between the SFHs of the ISLAndS dSphs and Milky Way
dSph companions of similar luminosities and host distances is the absence of 
very late quenching ($\tau_{q}$ $\le$ 5 Gyr) dSphs in the ISLAndS sample.
Thus, models that can reproduce satellite populations with 
and without late quenching satellites will be of extreme interest.

\end{abstract}


\section{INTRODUCTION}\label{secint}

\subsection{Motivation: Testing for Bias in the MW Satellites}

The nearby dwarf galaxies of the Local Group are unique probes of galaxy formation
and evolution over the entire history of the universe. Their proximity allows the 
study of their stellar, gaseous, and dark matter contents in unparalleled detail.
However, they are not pristine, primeval systems, and their evolution is dependent
upon both local and cosmic environmental factors. Thus, it is of tremendous
importance to disentangle these effects for nearby galaxies and
interpret them in a wider cosmological context.
In currently favored hierarchical structure formation models, density fluctuations on
the scale of dwarf galaxies collapse early and merge to form larger structures.
However, the accretion of gas and its conversion to stars in dwarf galaxies is
complicated and poorly understood, particularly at the earliest times.
Cosmological simulations predict vastly more surviving dwarf
galaxy sized halos than the number of observed dwarfs around the Milky Way (MW) and
M31 \citep[``the missing satellites problem,'' e.g.,][]{kwg93, kly_etal1999, 
moo_etal1999, bullock2010}.  The missing satellites problem cannot be solved by 
simply discovering more faint satellites to the Milky Way as emphasized in the
``too big to fail'' problem \citep{boylan2011, boylan2012} which is repeated again for M31
\citep{tollerud2014, collins2014}.
It seems clear that not all of these dark matter halos can retain baryons
and form stars. Processes such as cosmic reionization \citep[e.g.,][]{efstathiou1992,
bullock2000} and stellar feedback \citep[e.g.,][]{dekel1986,tassis2003} 
are invoked to suppress star formation or remove the gas from some subset of
dark matter halos.

Environmental effects are clearly important for the evolution of low-mass systems
in the Local Group. Gas-poor, pressure-supported dwarf spheroidal galaxies
(dSphs) are preferentially found as satellites of the MW and M31, whereas gas-rich,
rotating dwarf irregulars (dIrrs) are preferentially found in isolated locales
\citep[e.g.,][]{vdbergh1994a,grcevich2009}.
Additionally, the closest MW
dSph companions (distances $\le$ 100 kpc) have exclusively old stars with ages $\gtrsim$ 10
Gyr, while those more distant can show prominent young and
intermediate-age stellar populations \citep[e.g.,][]{vdbergh1994b, mateo1998, mcc2012, brown2014, weisz2014b}
and thus present a large variety of star formation histories (SFHs).
It is not clear whether this configuration is a generic outcome of hierarchical structure
formation models or a result of specific factors in the Milky Way's history.
Through detailed dynamical modeling, \citet{mayer2001a, mayer2001b, mayer2006} have shown that
``tidal stirring'' can remove most of the gas from a dwarf galaxy and transform
rotationally supported systems into pressure supported systems. However, the
existence of the isolated dSphs Cetus and Tucana, shown to be as old as the oldest
MW companions \citep{monelli2010b, monelli2010c}, point toward a multi-parameter process.

Very sophisticated models are being used to explore the environmental 
impacts on the evolution of dwarf galaxies \citep[e.g.,][]{ocvirk2014, onorbe2015, wetzel2015a,
benitez2016, sawala2016, wetzel2016}.
In addition, there is growing recognition that studying the nearest galaxies
provides an observational window on high redshift galaxy evolution that even the
next generation of high redshift galaxy surveys will not be able to provide
\cite[e.g.,][]{weisz2014d, boylan2015, patej2015, graus2016, boylan2016}.
Thus, it follows that obtaining observations of the nearest galaxies which
provide strong constraints on their lifetime SFHs is critical to our progress.  
  
Until very recently, the companions of the MW have been the only satellite dSphs
with robust derivations of their SFHs at {\it intermediate and old ages.}
The early SFHs of galaxies can only be revealed by observing resolved stars down to
and below the oldest main sequence turn-off (oMSTO) \citep[e.g.,][]{gallart2005}.  
Considering the unique role of the MW satellites as cosmological probes, it is vitally
important that we understand whether their early SFHs  are
representative of satellite dSphs in the wider universe.
The ISLAndS  project is the first opportunity to test the representative nature
of the early SFHs of the MW dSphs by obtaining complete, detailed SFHs for a representative 
sample of M31 satellites, the only other galaxy satellite system for which this is possible
with the presently available technology.
Our overall goal is to determine if the early evolution of the M31 companions is
significantly different from the MW companions, and, if so, to determine the local or
cosmic factors at play.  Thus, we can address the question: ``Are the
dSph companions to the MW truly representative of dSph galaxies in general?''

\subsection{M31 versus the Milky Way}

Is there any reason to suspect that the satellite populations of M31 and the MW 
could be significantly different?
There are significant differences between the properties of M31 and the MW
\citep{vdbergh1999} indicating
that their mass assembly histories were likely different.
M31 is generally assumed to be more massive, but the analysis of \citet{watkins2010} argues
that the two may have very similar halo masses.
M31 is thought to be an earlier type spiral, but \citet{beaton2007} revealed M31 to have
a boxy bulge, indicative of a bar, and making M31 a twin of the MW in that regard.
%The nuclear bulge of M31 is more luminous, and
%the velocity dispersion and the rotational velocity of M31 are larger.
% M31 has three times as many globular clusters [e.g.,][]
\citet{huxor2011} point out that M31 possesses a significant population of luminous 
and compact globular clusters (GCs) at large galactocentric radii without counterparts in the MW, and
that M31 also has a number of extended GCs, many of which are far larger than those 
in the MW.  They suggest that the differences between the two GC systems could be, 
at least partly, explained by the differing accretion histories that M31 and MW have experienced.
M31 appears to be the more massive and more evolved galaxy; yet,
recent accretion appears to be more important for M31 than
for the MW \citep[e.g.,][]{brown2006, bernard2012, deason2013, bernard2015a, bernard2015b, williams2015}.

\begin{figure*}[t]
\centering
\includegraphics[width=1.05\textwidth,angle=0]{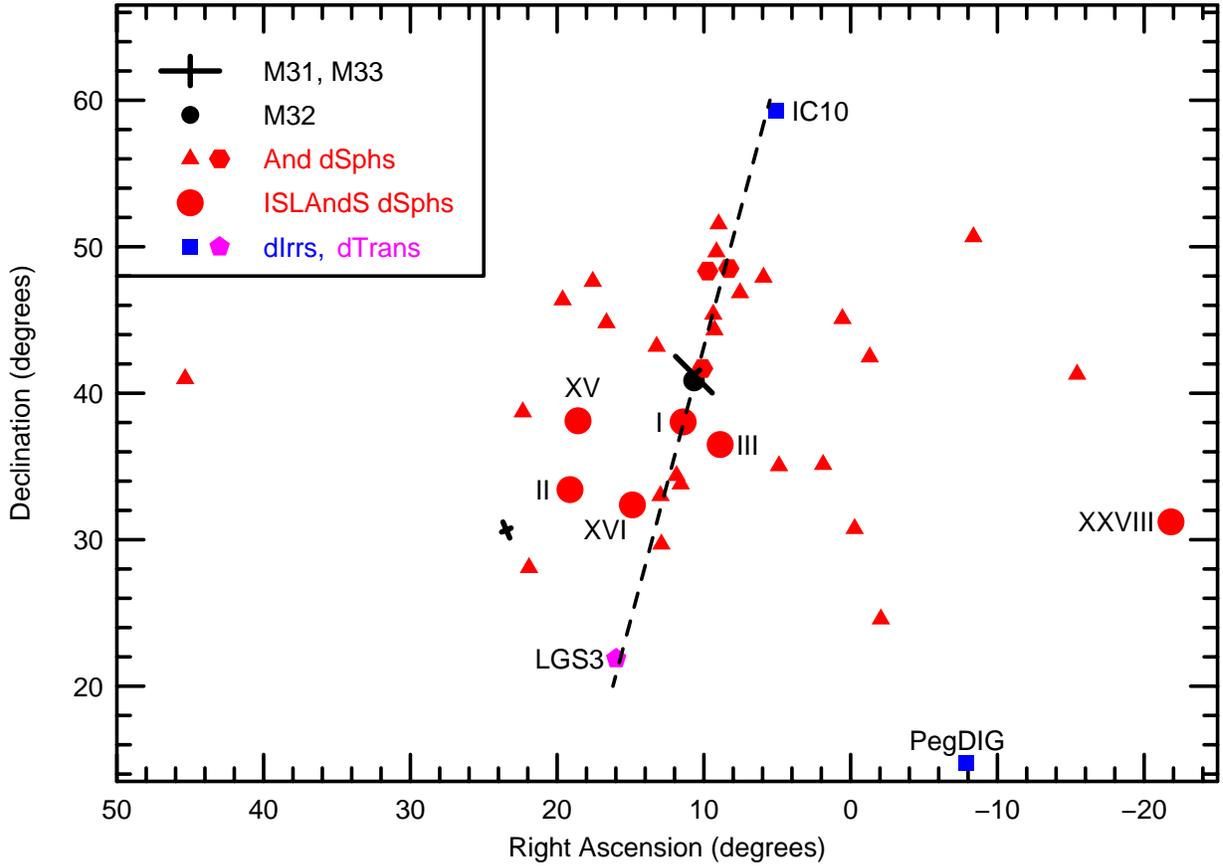}
\caption[ ]{The companions to Andromeda are plotted showing their
positions in the sky relative to M31 and M33.  The three high luminosity dSphs (NGC~147, NGC~185,
and NGC~205) are distinguished from the rest of the dSphs with hexagon symbols.
The positions of the ISLAndS sample are highlighted with larger symbols and labels.
Note that the ISLAndS sample galaxies span a large range in distance from M31.
The dashed line represents the approximate position of the thin plane discovered
by \citet{ibata2013}. Note that three galaxies from the ISLAndS sample
(And~I, And~III, and And~XVI) are located in the thin plane and three
(And~II, And~XV, and And~XXVIII) are outside of the plane.}
\label{f1}
\end{figure*}

The differences between M31 and the MW may extend to their
satellite populations.  
The presence of the true dE galaxy M32,
a relatively rare occurrence in nature \citep[see, e.g.,][]{kormendy2012}, indicates that
something special has taken place in Andromeda's satellite history, but exactly what remains
a topic for debate. The SFH for M32 derived by \citet{monachesi2012} shows a nearly constant
star formation rate up until 2 Gyr ago producing stars with nearly solar metallicities.

Another possible difference is the presence of the more luminous
dSphs NGC~147, NGC~185, and NGC 205. At M$_V$ $=$ $-$14.6, $-$14.8, and $-$16.5, respectively,
they are one to three magnitudes brighter than the MW's brightest dSphs, Fornax and Sagittarius, at
M$_V$ $=$ $-$13.4 and $-$13.5 \citep[][although as a tidally disrupting galaxy, the 
luminosity for Sagittarius may represent a lower limit]{mcc2012}.  This may be an indication of something significantly 
different in the formation of dSphs, or it may simply be the natural extension to
higher luminosities in a more abundant population.  \citet{geha2015} have produced SFHs
for NGC~147 and NGC~185 and found NGC~147 to have continued to produce stars well into 
intermediate ages while NGC~185 contains mostly older stars.  However, the direct interpretation
of these SFHs is complicated by the positions of the observed fields beyond the half-light 
radii; the extremities often show exclusively older stars even in 
actively star forming dwarfs.  So the
apparent surprise is the extended nature of the star formation in the outer regions of NGC~147.  
Note also that NGC~147 differs from NGC~185 in that NGC~147 shows the effects of a recent 
interaction \citep{crnojevic2014}.

Before this project, there were hints of possible differences between the M31 and MW dSphs.
For example, the M31 dSphs present redder horizontal branch
morphologies when compared to the MW dSphs \citep[e.g.,][]{dacosta1996, dacosta2000, 
dacosta2002, mcc2007}.
Additionally, \citet{mcc2006b} showed that the M31 dSphs
generally have larger half-light radii than the MW dSphs.
This was later quantified as  differences in the {\it mean} scaling relations
at the 1 - 2 sigma level \citep{brasseur2011, tollerud2012, collins2014}.
Regardless, the M31 dSphs occupy regions of parameter
space for which there are no analogous MW systems, emphasizing
that a full understanding of the origins of dwarf galaxy
properties cannot be obtained from the MW system alone.

\begin{figure*}[t]
\centering
\includegraphics[width=\textwidth,angle=0]{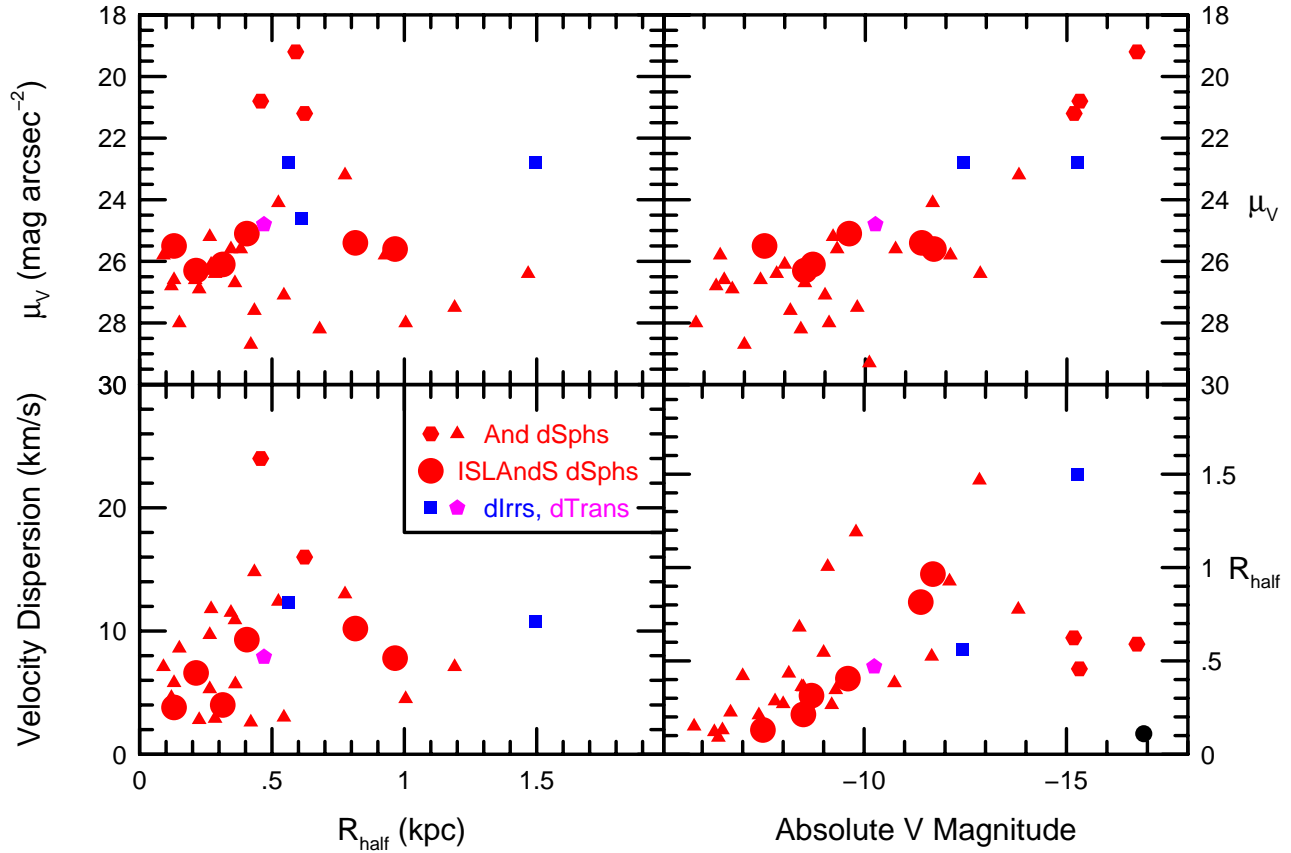}
\caption[ ]{Following \citet{kormendy1985} a diagram showing the intrinsic properties
(absolute V magnitude, half-light radius, central V surface brightness, and stellar 
velocity dispersion)
of the M31 satellites.  The symbols are identical to those used in Figure 1.  The distributions
of the ISLAndS sample relative to the rest of the M31 dSph companions show that the
ISLAndS sample is representative with the exception of the lack of very low luminosity
dSphs.  The lack of the lowest central surface brightness objects is a result of the 
lower limit on luminosity so that the number of observed stars would be large enough
to provide sufficient constraints on the history of the earliest star formation and the
correlation of central surface brightness with luminosity.  M32
does not appear in three of the panels due to its high central surface brightness (11) and
velocity dispersion (92).
}
\label{f2}
\end{figure*}

The early SFHs may possibly reflect the effects of the epoch of reionization.
The realization that the reionization of the universe
is quite inhomogeneous \citep[e.g.,][]{songaila2004,fan2006,becker2015} 
has led to a better appreciation of the impact
of the primary galaxy on the evolution of its satellites 
\citep[e.g.,][]{weinmann2007, busha2010, ocvirk2011}.
For example, simulations
by \citet{mayer2006, mayer2007} show that the local ionizing
radiation from the primary galaxy controls the temperature evolution and 
ionization state of the gas in the satellite dwarfs.
This sets the efficiency of mass stripping by tides and
gas removal by ram pressure from the dwarf satellites.
The {\it local} UV flux, at the distance of a typical MW satellite, 
is estimated to have been more than an order of magnitude higher
than the {\it average} cosmic UV background radiation at $z > 1$.
At that epoch, the primary galaxy was undergoing massive star formation
at levels comparable to present-day starburst galaxies \citep{governato2007}.
Since the intensity and temporal evolution of the radiation field
of the primary galaxy will depend on its SFH and mass assembly history,
given the differences between the MW and M31, we might expect the evolution
of their satellite galaxies to have been significantly different.

The structure of the paper is as follows: 
The ISLAndS sample is described in \S\ref{sample}.
The observations and data reduction are presented in \S\ref{secred}. 
The SFHs of the sample galaxies are presented in \S\ref{secsfhresult}.
A comparison of the quenching times for the sample galaxies is 
given in \S\ref{secquench}.
A comparison of the properties of the sample galaxies within and outside of the
thin plane identified by \citet{ibata2013} is presented in \S\ref{secplane}.
Finally, we make our first attempts at our main goal, comparing the M31 and
Milky Way satellites in \S\ref{seccomp}. 
The main conclusions of the work are summarized in \S\ref{seccon}.
In this work, cosmological parameters of 
$H_0=70.5\rm~km~s^{-1}~Mpc^{-1}$, $\Omega_m=0.273$, and a
flat Universe with $\Omega_\Lambda = 1 - \Omega_m$ are assumed
\citep[i.e.,][]{kom_etal2009}.

\section{The ISLAndS Representative Sample}\label{sample}

\subsection{Properties of the ISLAndS Galaxies}

M31 has a diverse satellite galaxy population \citep[e.g.,][]{mcc2006a},
and the number continues to grow with the discovery of increasingly fainter galaxies 
\citep[e.g.,][]{mcc2008, richardson2011, slater2011, bell2011, martin2013a, martin2013b}.
For the ISLAndS program we are focusing on the M31 dSphs more luminous than M$_V$~$\lesssim$~$-$7
in order to derive well-constrained SFHs for comparison to their MW analogues.
These galaxies are massive enough that their status as galaxies is
not controversial, and they are populous enough
to provide strong constraints on their SFHs over the age of the universe.
The abundance of M31 companions allows us
to design a representative sample - which spans the range of properties of the
ensemble - yet consists of the galaxies which are least expensive to observe.

Our sample of six galaxies is presented in Table~1, and has
been selected by balancing exposure time considerations with the requirement
to observe galaxies spanning a range of luminosity (M$_V$), half-light radius
(R$_{H}$), and distance from M31 (D$_{M31}$).  Thus, the sample consists of 
galaxies with minimal distances from us and minimal foreground extinction. 
Because of the large angular size of the M31 satellite distribution and the 
relatively low Galactic latitude of M31 (b $=$ $-$21.6$\degr$) there is a large range
in foreground reddening to the satellites 
\citep[0.04 $\le$ E(B-V) $\le$ 0.20][]{mcc2012}.  Thus, there is a significant
range in required exposure times for the M31 satellites to be observed to the 
required depth.

The positions in the sky of the ISLAndS sample are shown in Figure \ref{f1}.
Figure \ref{f1} immediately shows that the ISLAndS sample spans a large
range in projected distance and the true distances from M31 range 
from 58 to 370 kpc (Table~1).
Based on ``fragmentary'' data, \citet{vdbergh1994b} first pointed out the general trend
for the stellar content of faint Local Group dwarfs to correlate with distance from the
Galaxy.  This correlation is in the sense that the closest have uniformly old stellar
populations while the more distant have larger intermediate age populations.  While not
without exceptions, modern observations have shown that this trend is still valid.
A goal of our program is to determine whether the M31 dSphs show a similar trend of
increasing mean age with distance from host as shown by the Galactic dSphs.

In Figure \ref{f2} we show a ``Kormendy'' diagram \citep{kormendy1985} 
presenting the intrinsic properties of all of the M31 satellites and 
highlighting the ISLAndS sample.  This figure was assembled from data in \citet{mcc2012},
updated with data from \citet{conn2012} and Martin et al.\ (2016), and the data in
Table~1.  As can be seen in Figure \ref{f2}, the ISLAndS
sample galaxies give a good representative coverage of the range of intrinsic
properties of the M31 dSph satellites.  The exception is a lack of 
very low surface brightness galaxies in the ISLAndS sample.  This is a result of
the lower limit on luminosity imposed in order to obtain strong constraints on 
the SFHs.

\begin{figure}[h]
\centering
\includegraphics[width=8.5cm]{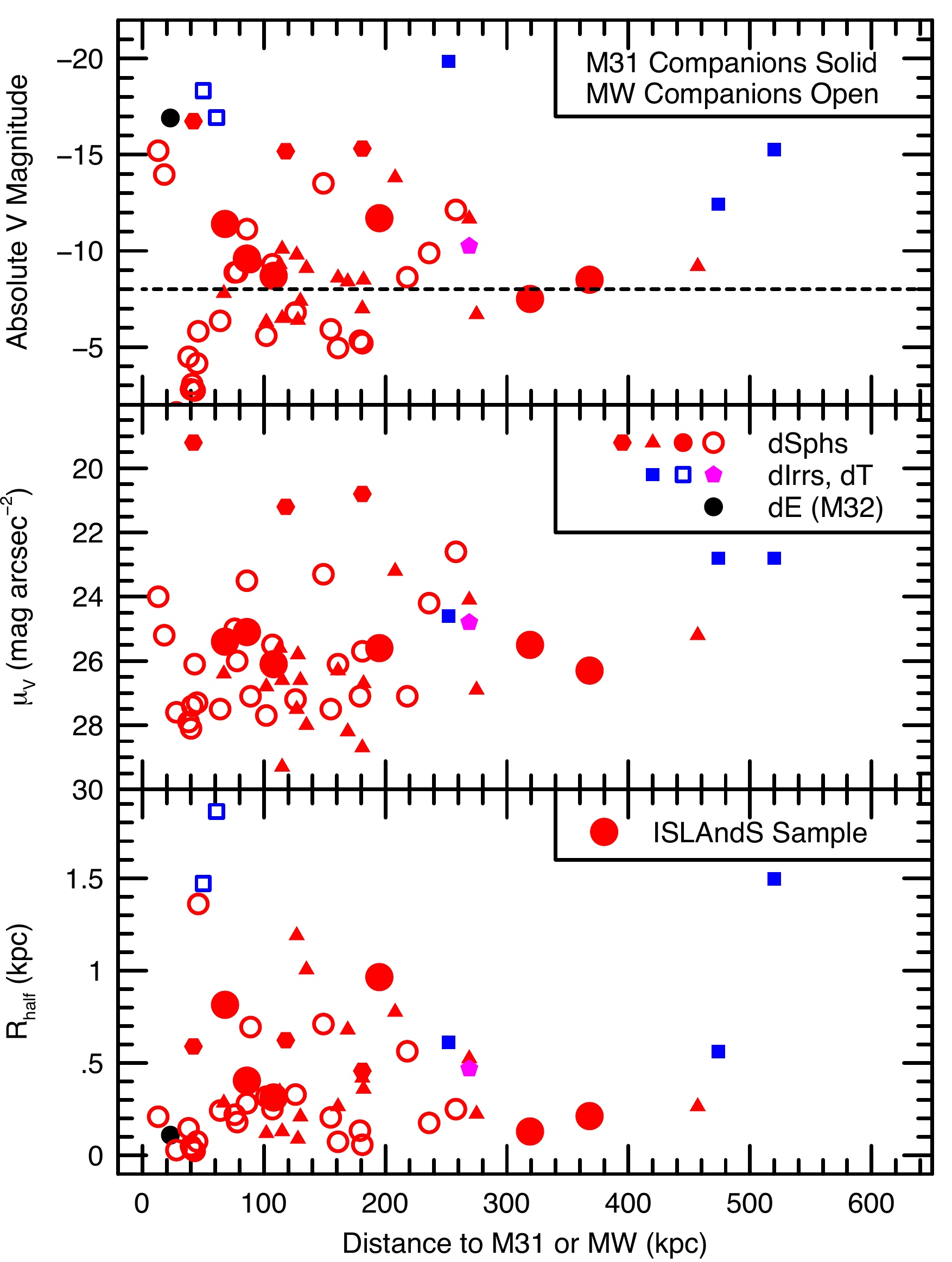}
\caption[ ]{The properties of companions to Andromeda (luminosity, central surface brightness,
and half-light radius) as a function of the distance
to Andromeda highlighting the positions of the ISLAndS sample observed here.
For comparison, the satellites of the Milky Way are added.
The dashed line in the upper plot indicates the lower limit for the ISLAndS sample to 
provide sufficiently populous color magnitude diagrams for strong constraints on the SFHs.
Note that M32 is well off the scale of the middle panel of this plot
due to its high central surface brightness of 11.1. 
}
\label{f3}
\end{figure}

In Figure \ref{f3}, we have plotted a comparison of the half-light radii, central surface
brightnesses, and luminosities of the M31 and MW satellites as
a function of distance from the host galaxy.
Figure \ref{f3} shows that the ISLAndS M31 dSphs span similar ranges in luminosity
and radial distance from host as the well-studied MW dSphs.
These true three-dimensional separations
are based on differential heliocentric distances
\citep[see:][]{mcc2004, mcc2005, mcc2006a, mcc2012, conn2012}. 

Phase~I of this project (HST cycle 20, observed November 2013) started with
observations of And~II and And~XVI \citep[see][]{weisz2014a}.  Cycle 20 Hubble 
Space Telescope proposals for targets in the restricted 
RA zone around Andromeda were limited to a total of 30 orbits, so we
proposed the two galaxies that could be done in 30 orbits.
Our cycle 22 program allowed us   
to extend our radial coverage to smaller (And~I and And~III)
and larger radii (And~XXVIII) allowing a direct comparison of inner vs.\ outer
galaxies and also to fill in the middle in the M$_V$, R$_H$ plane.  
And~XXVIII and And~XV also address
the critical question of whether And~XVI is an anomaly or shows the importance
of separation distance over mass for early quenching 
\citep[see discussions in][]{weisz2014a, monelli2016}.  
Together these six galaxies 
allow us to test the hypothesis that differences between MW and M31 dSphs
are due to early evolution of {\it the parent galaxy}.  Additionally, half
of the sample are in the thin plane of co-rotating galaxies identified by \citet{ibata2013}.

\subsection{Kinematic and Abundance Data from the Literature}

In Table~2, we have assembled data from the literature for the kinematics and 
chemical abundances for the ISLAndS sample galaxies.  These data come from spectroscopic
studies of individual RGB stars in these galaxies and represent a significant
investment of ground-based (especially Keck) observing time.  We have two reasons 
for the compilation in Table~2.  First, we would like to present the original sources
for reference for the masses and chemical abundances for discussion of the ISLAndS galaxies.
Secondly, while there has been tremendous progress on this front, we would like to highlight
that more work is still needed.  In many regards, And~II is the ideal example of what
can be learned.   With hundreds of spectra observed, the velocity dispersion is very
well defined, in fact, the large number of spectra allowed \citet{amorisco2014} to identify
a kinematically cold component in And~II.  Note that there is no overlap with our HST
fields of view and the kinematically cold component in And~II, and, 
to date, this component has not been observed
with the HST in order to study its SFH.    

Spectroscopic chemical abundances also hold promise for a better understanding of the 
evolution of these galaxies.  
Spectroscopic stellar abundances provide a way to increase the precision in determining 
the star formation histories at early ages where the color-magnitude diagram 
technique has challenging time resolution limitations \citep{deboer2012a, deboer2012b, 
deboer2014, brown2014, dolphin2016}.  The improved early time resolution of these star formation
histories arises because the spectroscopic abundances constrain potential degeneracies
between age and metallicity.  
However, some care needs to be exercised, as, to date, these techniques have not been demonstrated
using multiple stellar libraries.  Since the dominant uncertainty in deriving 
star formation histories is the systematic uncertainty of choosing a stellar evolution library
(see Appendix),
we regard these improvements in time resolution with a degree of skepticism
\citep[see discussion in][]{dolphin2016}.  
Regardless, ambitious spectroscopic abundance studies primarily hold the promise for significantly
more reliable age-metallicity relationships (AMRs) and these are vital to a complete
understanding of the evolution of these galaxies \citep{dolphin2016}.  Additionally, as pointed out in 
\citet{mcc2010}, multiple epochs are highly desired to eliminate inflated velocity
dispersions due to binaries. 

Given the value of the investment of HST observing time dedicated to the study of the
ISLAndS sample, we heartily encourage additional ground-based spectroscopic campaigns 
to bring the other galaxies up to the remarkable standard achieved for And~II.  We note
especially the near absence of relative chemical abundances (e.g., [$\alpha$/Fe]).  
\citet{dotter2007} has emphasized the important role that accurate relative chemical
abundances can play in deriving SFHs.

\begin{figure*}[h]
\centering
\includegraphics[width=0.85\textwidth,angle=0]{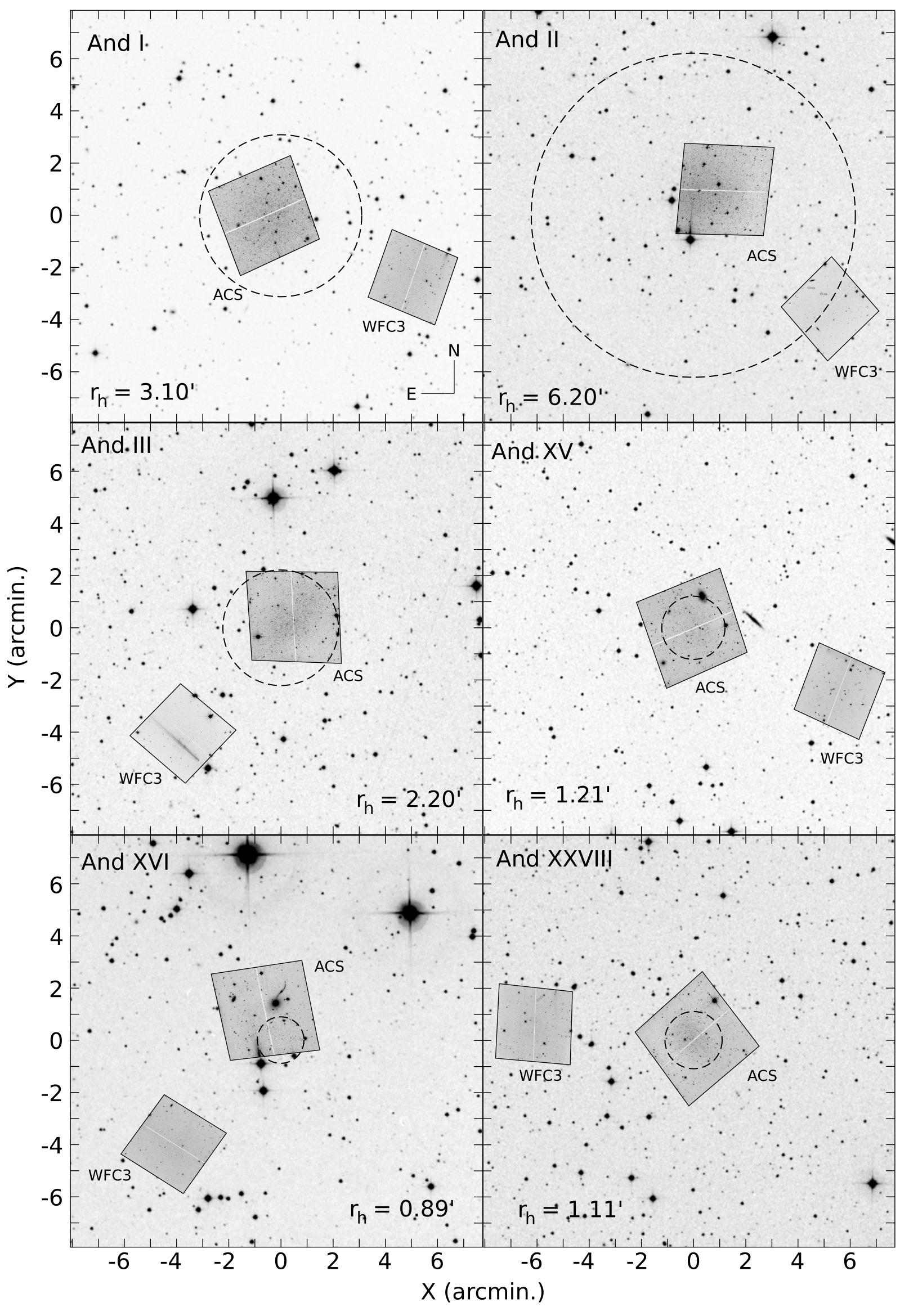}
\caption[ ]{The HST/ACS observing positions for the ACS and WFC3 cameras for
the ISLAndS sample.  The dashed circles show the half light radii for the
galaxies. The ACS field for And~XVI was not centered in order to avoid a bright
foreground star.  The ACS fields for And~I, And~II, and And~III were offset from 
the galaxy center to provide a better radial coverage.
}
\label{f4}
\end{figure*}

\section{OBSERVATIONS AND DATA REDUCTION}\label{secred}

\subsection{The Observations}

The HST observations of the six M31 satellites were obtained between October 2013 and
September 2015.  The two galaxies observed in cycle 20 (And~II and And~XVI) have  
previously been reported in \citet{weisz2014a}.  All observations were reduced in 
a uniform manner, and a brief summary is provided here.

Following the observing protocols established by the LCID program
\citep[e.g.,][]{monelli2010b, monelli2010c, hidalgo2011}, the F475W and F814W bands 
were selected as the most efficient combination to trace
age differences at old ages, since they provide the
smallest relative error in age and metallicity in the main-sequence 
and sub-giant regions.  Asymmetric exposure times were chosen such that
the uncertainties in the two different photometry bands were equal at
an F475W-F814W color of 1 -- essentially the color of the main sequence one 
magnitude below the oMSTO.

The observations were organized into two orbit visits,
and each orbit was split into one F475W and one F814W exposure (in 
order to maximize sampling of variable star light curves).
Each visit acquired 2363s of integration time in F475W and 2088s of
integration time in F814W with the ACS. The total integration times are given
in columns 3 and 4 of Table~1.  
Simultaneously, parallel fields (see Figure~\ref{f4}) were observed with the
WFC3 camera with exposure times of 2759s in F475W and 2322s in F814W. 
Dithers of a few pixels between exposures were introduced to
minimize the impact of pixel-to-pixel sensitivity variations (``hot pixels'') 
in the CCDs. As with the LCID program, the visits were planned to take place over 
several days in order to properly sample the light curves for variable stars with 
periods less than $\sim$ two days.  Since it was anticipated that the bulk of the 
variable stars would be RR Lyrae with periods of roughly 0.5 days, the scheduling
of the visits was designed to minimize cadences of 12 hours ($\sim$ 8 orbits) so
that the observations would not all be taken near the same phase.  This worked
well for the LCID program \citep[e.g.,][]{bernard2009, bernard2010} and also 
for the present program.  

The positions of the observed fields are shown in Figure \ref{f4}.  
As can be seen in Figure \ref{f4}, for And~I and And~II the ACS
field covers only part of the galaxy within the half light radius and the
parallel WFC3 field contains many member stars.  
Some member stars were detected in the WFC3 field of And~III, but not
nearly enough to produce a SFH with reasonable uncertainties.  
For And~XV, And~XVI, and And~XXVIII, the ACS field of view covers 
most of the galaxy out to the half light radius and the WFC3 field is distant 
enough that a minimal number of member stars are expected.

\subsection{Data Reduction}

We analyzed images taken directly from the STScI pipeline (bias, flat-field,
and image distortion corrected) working with the charge transfer efficiency 
corrected images (i.e., .flc images). Two PSF-fitting photometry packages,
DAOPHOT/ALLFRAME \citep{ste1994} and DOLPHOT, an updated version of HSTPHOT with 
ACS and WFC3 specific modules \citep{dol2000},  
were used independently to obtain the photometry of the resolved stars.
See \citet{monelli2010b} for more details about both photometry reduction
procedures.  
Individual photometry catalogs were calibrated using equations provided 
by the STScI \citep[e.g.,][]{sir_etal2005} with the most recent updates 
(e.g., ACS ISR 12-01).  
The differences between the two sets 
of photometry are small and typical for obtaining HST photometry with different
methods \citep{hill98, holtzman06}, and so, for simplicity, the rest of this 
paper is based on only the DOLPHOT photometry datasets.  
%
%The DOLPHOT photometry
%followed the optimized prescriptions identified by the PHAT project 
%\citep{williams2014} and uses the Anderson (ACS ISR 2006-01) PSF library. 
%
%From the raw photometric catalog, we rejected objects that did not meet 
%particular requirements in signal-to-noise (S/N), PSF profile sharpness, 
%and whose flux was significantly affected by neighboring objects. Specifically, 
%our accepted stars have S/N$_{F475W}$ and S/N$_{F814W}$ $>$ 5, 
%(sharp$_{F475W}$ + sharp$_{F814W}$)$^2$ $<$ 0.1, 
%and (crowd$_{F475W}$ + crowd$_{F814W}$) $<$ 1.0. 
%The precise definitions of these criteria can be found in Dolphin (2000).

Signal-to-noise limitations, detector defects, and stellar crowding can all
impact the quality of the photometry of resolved stars with the resulting
loss of stars, changes in measured stellar colors and magnitudes, and
systematic uncertainties. To characterize these observational effects, we injected
$\sim 10^6$ artificial stars in the observed
images and obtained their photometry in an identical manner as for the real stars.
\citet{monelli2010b} and \citet{hidalgo2011} provide detailed descriptions of the procedures
we adopt for the characterization and simulation of these observational effects.

\begin{figure*}[h]
\centering
\includegraphics[width=\textwidth,angle=0]{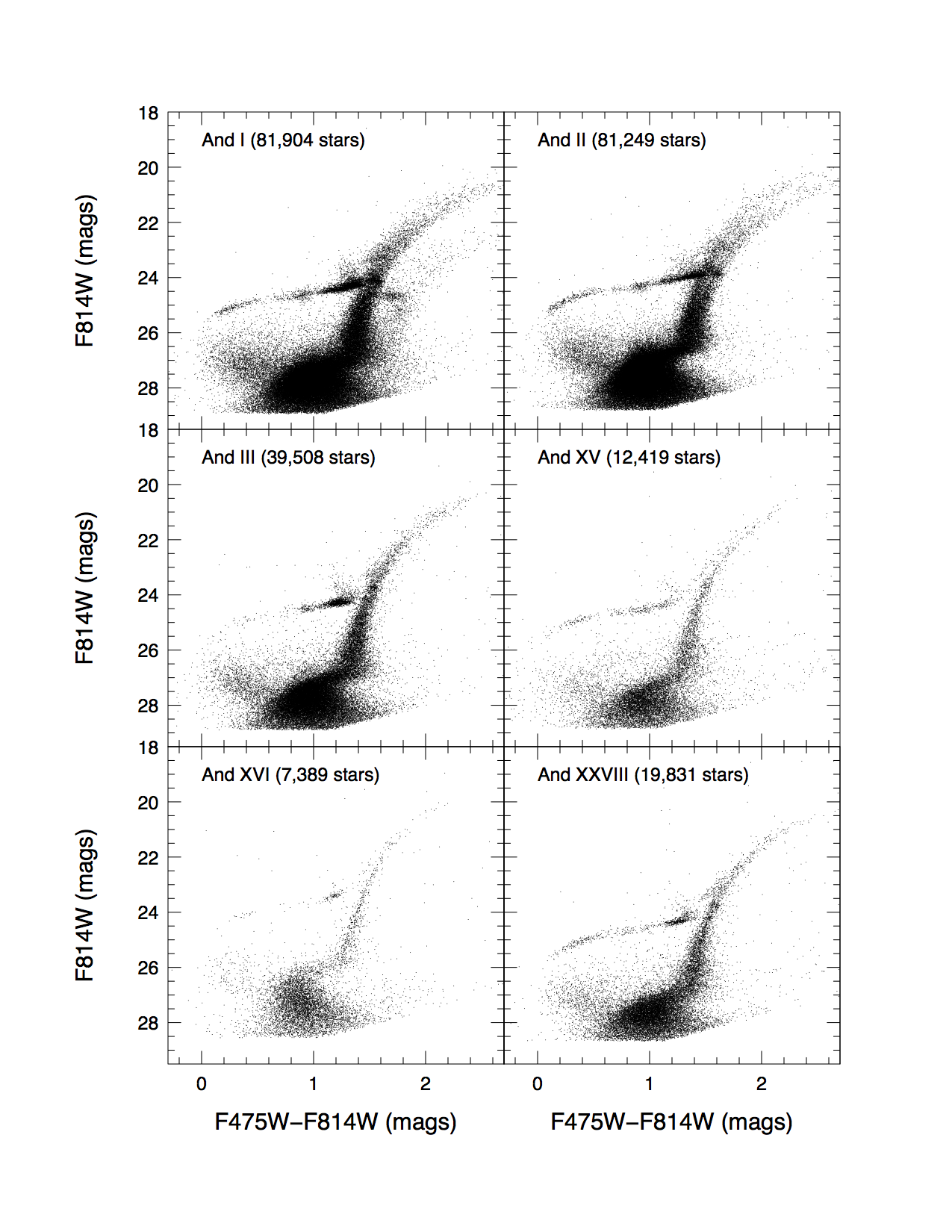}
\caption[ ]{The HST/ACS color-magnitude diagrams of the ISLAndS sample.
The number of sources which passed the quality cuts are reported for
each galaxy.  In all cases, the photometry reaches to below the oldest
main sequence turnoffs.  The CMD for And~I shows contamination from 
Andromeda's Giant Stellar Stream \citep{ibata2001, ferguson2002, mcc2003}.
The extra depth in And~XVI is due to a
preliminary distance measurement that was over-estimated.
}
\label{f5}
\end{figure*}

\subsection{The Color Magnitude Diagrams}

\subsubsection{The ACS CMDs}

The CMDs for the ACS fields of the observed M31 satellites are shown in Figure \ref{f5}. 
The left axis shows the observed F814W magnitudes in the ACS photometric system uncorrected for extinction. 
Figure \ref{f5} shows that the ACS photometry reaches below the 
oldest main-sequence turn-offs
% at the 50\% completeness limit,
allowing for very strong constraints on the oldest epochs of star formation.
These observations of And~I, And~II, and And~III are $\sim 2.5$ mag fainter 
than the deepest CMDs previously obtained for these galaxies \citep{dacosta1996,
dacosta2000, dacosta2002}.

There are several notable features that are common to all six ACS CMDs.
All six galaxies show the presence of blue horizontal branch stars (BHBs).
This is generally taken as evidence of the presence of stars with ages comparable to
the Milky Way GC population \citep[see discussion in][]{gallart2005}.  
To date, all galaxies that have been observed with sufficient photometric depth have 
shown at least some star formation at the earliest times.
\citet{dacosta1996, dacosta2000, dacosta2002} noted that And~I, And~II, and And~III
all had substantial populations of red horizontal branch stars (RHBs) and this
was interpreted to mean that all three galaxies had substantial populations of
stars younger than the typical Milky Way GCs.  The connection between RHBs and
intermediate age stars is generally sound. However, a direct connection between
HB morphology and SFH remains elusive.  For example, the Milky Way companions
Draco and Ursa Minor have identical absolute magnitudes and very different
horizontal branch morphologies.  Draco has a predominantly red HB and Ursa Minor
has a predominantly blue HB, yet SFHs derived from oMSTO photometry seem to 
indicate that Draco 
formed most of its stars before Ursa Minor \citep{olszewski1985, grillmair1998,
mighell1999, weisz2014b}.  
Thus, while the HB morphology 
can provide clues to the old age SFHs, there is no substitute for oMSTO 
photometry.

In fact, all six galaxies show complex horizontal branch (HB) morphologies, indicative 
of a metallicity or age range. In all six galaxies, the HB is extended from the 
blue to the red side, with the population of RR Lyrae variable stars noticeable 
as a clump at (F475W-F814W) $\sim$ 0.9 (in some cases spanning a wider magnitude 
range than the HB stars at either side). The relative populations of the blue and 
red HBs varies between galaxies, with And~I, And~II, and And~XXVIII having both 
the blue and the red HBs well populated, while And~III and And~XVI have more 
prominent red HBs. In And~I and And~II the red HBs even merge with the RGBs, 
which are very wide, indicating a wide metallicity range.

\begin{figure*}[t]
\centering
\includegraphics[width=\textwidth,angle=0]{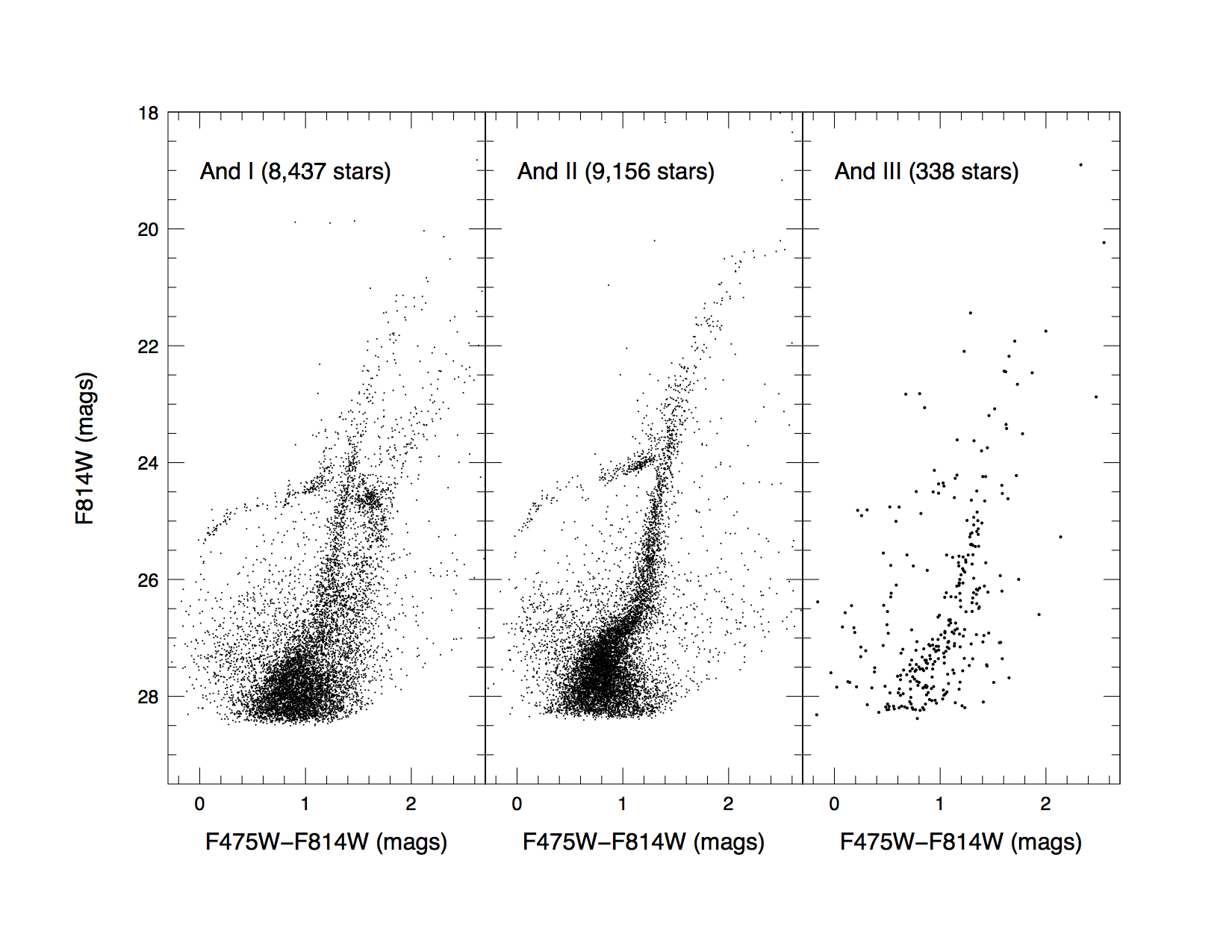}
\caption[ ]{The HST/WFC3 color-magnitude diagrams of the ISLAndS sample.
The number of sources which passed the quality cuts are reported for
each galaxy.  In both cases, the photometry reaches to the oldest
main sequence turnoffs, but is not as deep as the ACS photometry.
Because of its larger radius from the galaxy, the WFC3 CMD for And~I is
dominated by contamination from Andromeda's Giant Stellar Stream 
\citep{ibata2001, ferguson2002, mcc2003}.
}
\label{f6}
\end{figure*}

All six galaxies also show a significant blue plume in the CMDs 
below the horizontal branch and above the oldest MS turnoffs  
which is typically associated with a ``blue straggler'' population. 
If these stars are interpreted
as MS stars, they have ages down to $\sim$ 2 Gyr.  However, these populations are
ubiquitous in dSph galaxies (at low levels), and, there are good arguments that they are due
to the altered evolution of coalescing primordial binary stars 
\citep[see discussions in][]{momany2007, mapelli2007, mapelli2009, monelli2012,
santana2013}.  Note also the recent work by \citet{gosnell2014, gosnell2015} finding 
direct evidence of mass transfer in the blue straggler star population of the 
MW open cluster NGC~188.
However, to date, no one-to-one relationship has been proposed for the ratio
of old age stellar populations to blue straggler stars, so 
the possibility exists that in some dSph galaxies there could be younger MS
stars mixed in with (or, equivalently, hidden by) the blue straggler population.

All six galaxies also show, to differing degrees, the presence of an upward extension 
to the red clump.  In models, this part of the
color magnitude diagram is populated by helium burning stars with ages between
$\sim$ 2 - 4 Gyr.  Helium burning stars with even younger ages can also occupy this
region, but they would be accompanied by a significant population of bright MS 
stars which we do not see in the present CMDs.  Note that in the blue straggler 
interpretation, there is an upper limit to the MS stellar mass at twice the mass
of the stars at the turnoff corresponding to the quenching time.
At a constant metallicity,
the positions of these intermediate age red clump stars form a spur in the
CMD, which initially increases in luminosity and color (getting redder) with 
decreasing age, and then continues to increase in luminosity but turns bluer
with even younger ages.  These spurs off of the red clump are displayed prominently in the
CMDs of Leo~A \citep{cole2007}, IC~1613 \citep{skillman2014}, and 
Aquarius \citep[DDO~210;][]{cole2014} because of the continuous star formation
in these galaxies.  The presence of these spurs in the ISLAndS dSphs is likely a reflection
of the blue straggler population already noted.  The detailed modeling presented
in the next section shows that the ratio of blue straggler stars to these stars is 
consistent with this scenario.

There is one feature that is lacking in all six galaxies, and that is obvious,
distinct multiple generations of star formation as evidenced by multiple sequences in 
the subgiant region.   This is not a common feature of dSph galaxies, but it
is seen famously in the Milky Way companion Carina \citep{monelli2003, bono2010,
weisz2014b}.  In this regard, the SFH of Carina remains unique.

There are other features of the individual galaxy CMDs worth noting.  The CMD
of And~I is contaminated by Andromeda's Giant Stellar Stream \citep{ibata2001,
ferguson2002, mcc2003}.  This stream is made up of relatively metal rich stars
and is consistent with a relatively recent tidal origin, now thought to be 
the result of the complete disruption of an infalling satellite by M31
\citep{ibata2004, fardal2013, sadoun2014}.
%likely due to a prolonged
%episode of tidal stripping from Andromeda's nearest neighbor satellite M32.
This stream is located $\sim$100 kpc behind And~I, and thus, features are easily 
visually separated from And~I in the CMD.   The stream's RGB appears redder and 
fainter than that of And~I and has a prominent red clump also redder than the
RGB of And~I.

The bifurcated RGB in And~II is another notable feature in Figure \ref{f5}.
This feature was first noted by \citet{dacosta2000} and attributed to distinct
populations in metallicity.  \citet{mcc2007} noted that the stars of the red
branch of the RGB were centrally located, adding further evidence for two
distinct populations.  \citet{weisz2014a} have previously presented this
CMD of And~II and work is in progress to combine wide-field ground-based 
imaging along with our ACS and WFC3 observations to explore spatial variations 
in the populations of And~II. 
 
The photometry of And~XVI is deeper than those of the other galaxies.  This is 
because when the observations were taken, the distance to And~XVI was thought
to be slightly larger than the accepted value today.  As a result, the 
luminosity of And~XVI is less than originally thought and the CMD is 
correspondingly less populated than originally anticipated.   

\subsection{The WFC3 Parallel Fields}

For all six ISLAndS galaxies, parallel observations were obtained with the 
WFC3 camera in F475W and F814W (see Figure~\ref{f4}).  In three of the
galaxies (And~XV, And~XVI, and And~XXVIII), there was essentially no
detectable stellar population.   

%there were insufficient 
%galaxy stars detected for construction of a SFH, and thus, these will not be discussed
%further.  For the two galaxies with significant detections in the parallel
%field, the total integration
%times were 29,667s in F475W and 23,639s in F814W in And~I and 23,482s in 
%F475W and 19,776s in F814W in And~II.

Figure \ref{f6} shows the CMDs for the WFC3 parallel fields in And~I, And~II, and And~III.
In And~I, the WFC3 CMD appears to be very similar to that of the ACS except that 
it is much less populous.  The giant stellar stream is more prominent relative
to And~I, as would be expected.  In And~II, only the older, metal poor RGB is
present (brighter and bluer); there is no evidence of the second RGB, which 
indicates that the younger, more metal rich stars are more centrally 
concentrated.  In And~III the stellar population is detectable, but there are
insufficient stars for strong constraints on the SFH. 
The SFHs of the outer fields will not be presented here.
These fields will be used for detailed studies of the radial gradients in
stellar populations for these galaxies.  The WFC3 CMDs are presented here 
for completeness.

\section{THE SFHs of M31 Satellites}\label{secsfhresult}
\subsection{Distances}

The most populous CMDs allow distance determinations from the tip of the RGB (TRGB).  
Following \citet{makarov2006}, we used a 
%Bayesian 
maximum likelihood method
to determine the magnitude of the RGB tip.  Four of the galaxies returned
secure estimates and these are listed in Table~3. The TRGB magnitudes were converted
to distances assuming the ACS F814W calibration from \citet{rizzi2007} of
M$^{TRGB}_{I}$ $=$ $-$4.05.
In two cases (And~XVI and
And~XXVIII), the maximum likelihood code returned ambiguous answers, so we adopted 
distances resulting from the best SFH solutions (described later in this section).

The derived TRGB distances are in relatively good agreement with the TRGB
distances from \citet{conn2012}.  In general, our new TRGB distance moduli average
~0.1 magnitudes larger.  The one exception is And~XV where our distance modulus
(24.66) is significantly larger than the 23.98$^{+0.26}_{-0.12}$ reported by
\citet{conn2012}.  We are also deriving distances using the RR~Lyrae populations
in the galaxies (Mart{\'i}nez-V{\'a}zquez et al.\ in prep.) and these support the 
longer distance estimate.  \citet{letarte2009} noted that the brightest three stars 
which had previously been used to define the tip of the red giant branch in And~XV were
foreground Galactic stars, and derived a distance modulus of 24.43.
Note that systematic uncertainties in distance
estimates can play a significant role in the determination of the star formation 
rates at the earliest times (ages $\ge$ 5 Gyr) where a $\sim$0.1 magnitude difference in MSTO 
luminosity corresponds to a $\sim$1 Gyr difference in age.  For a reliable 
comparison of the earliest SFHs of MW and M31 satellites, a uniform distance
measurement methodology will need to be adopted.

To determine the absolute magnitudes in Table~3, we used the photometry from
Martin et al.\ (2016), except for And~XXVIII which was not included in
their study.  For And~XXVIII, we use the apparent magnitude from \citet{slater2011}.
A comparison between the apparent magnitudes for the M31 satellites in \citet{mcc2012}
and those in Martin et al.\ revealed a significant scatter with a
systematic trend for fainter apparent magnitudes in the Martin et al.\
photometry for the more luminous galaxies.  Typically, the differences in 
apparent magnitude are larger than the quoted uncertainties.  In an attempt to have all 
six galaxies on the same absolute magnitude scale, we made an adjustment of 0.5 
magnitudes to the apparent magnitude of And~XXVIII from \citet{mcc2012}.  For comparison,
in the final
column of Table~3, we add the  
total mass within the half-light radius calculated from the half-light radius from Table~1 
and the stellar velocity dispersions
(latest values from Table~2) and using the mass estimator from \citet{walker2009}.

%To give an overview of the lifetime SFHs of the ISLAndS sample galaxies, 
%in Figure (TBD), we
%present 3-D representations of their SFRs and metallicities as a function 
%of time 
%\citep[``population boxes'', following][]{hodge1989}.

\subsection{Methodology}

These SFHs were derived using two parallel methods.  One is known as the
IAC method, consisting of
IAC-star \citep{apa_gal2004}, IAC-pop \citep{apa_hid2009}, and MinnIAC
\citep{apa_hid2009, hidalgo2011}.  These codes allow simultaneous solutions 
for the SFH and the age-metallicity relationship (AMR) with total uncertainties
estimated from contributions from statistical, binning, distance, and reddening 
uncertainties.
Details of the methodology can be found in \citet{hidalgo2011}.
Specifically, solutions were obtained using the BaSTI
\citep{pie_etal2004} stellar evolutionary libraries and the bolometric
corrections of \citet{bed_etal2005}. 
%The uncertainties are the
%statistical uncertainties and do not include certain systematics (e.g.,
%the use of a different stellar library).

The other code is MATCH \citep{dol2002}.  MATCH has been recently modified
to calculate systematic uncertainties associated with the choice of
stellar evolution library \citet{dolphin2012} and statistical uncertainties 
\citep{dolphin2013}.
The systematic uncertainties are estimated through application of shifts in 
temperature and luminosity which mimic the differences between stellar evolution libraries.
For the statistical uncertainties, random uncertainties were generated using 
the Hybrid Monte Carlo algorithm of \citet{duane1987}.  
The result of this Markov Chain Monte Carlo routine is a sample of 10,000 SFHs
whose density is proportional to the probability density.  
Upper and lower random error bars
for any given value (e.g., cumulative stellar mass fraction at a particular point in time) are
calculated by identifying the boundaries of the highest-density region containing 68\%
of the samples, with the value 68\% adopted as it is the percentage of a normal
distribution falling between the $\pm$ 1 $\sigma$ bounds. 

The results from the two different codes were compared and found to be in
agreement within the uncertainties (when using identical stellar evolution libraries).  
Here we present the results based on the MATCH code.
These will allow a more uniform comparison to previous studies of other galaxies
in the literature.

\begin{figure}[h]
\includegraphics[width=9cm,angle=0]{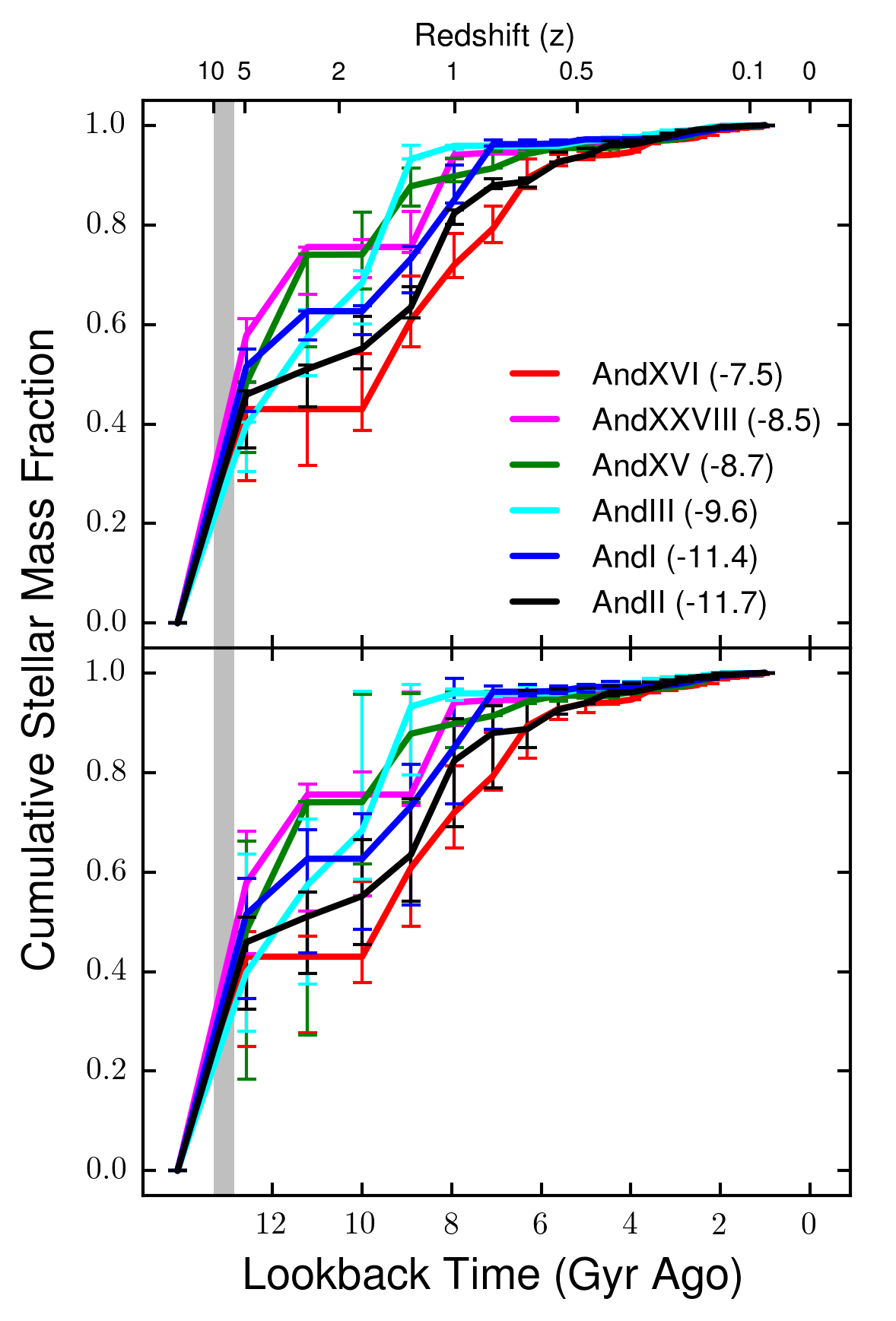}
\caption[ ]{
Comparison between the SFHs of the ISLAndS galaxies shown as cumulative
stellar mass fraction.
The upper panel shows only the statistical uncertainties,
and the lower panel additionally accounts for estimated systematic uncertainties as discussed in
\citet{dolphin2012, dolphin2013}.
The redshift scale given on the top axis is
calculated assuming a concordance $\Lambda$CDM cosmology.
The shaded portion of the graph shows the era of cosmological reionization from
$z$ $\approx$ 10-6.
The galaxy absolute V magnitudes are shown in the legend.
}
\label{f7}
\end{figure}

\subsection{The SFHs}

We next present a quantitative comparison of the lifetime SFHs for the 
six M31 dSphs in the ISLAndS sample.
As discussed in the Appendix, the dominant source of systematic uncertainty in the
determination of a SFH from a CMD (of a given photometric depth) is the choice of 
the stellar evolution library.  Thus, when comparing SFHs of different galaxies, it is 
important to use the same library (and, ideally, to conduct the comparison 
multiple times with different libraries).  For a first look at the 
SFHs of the ISLAndS sample, we use the MATCH code  and the Padua
\citep{girardi2010} stellar evolution library.  We choose the Padua stellar evolution
library to allow for direct comparisons with previous work \citep[e.g.,][]{weisz2014b}.

In Figure \ref{f7}, we show a comparison between the SFHs of the six ISLAndS galaxies as
cumulative stellar mass fractions.  In the upper panel, only the statistical
uncertainties \citep[as per][]{dolphin2013} are shown.  
In the lower panel, the combined statistical and systematic uncertainties 
\citep[as per][]{dolphin2012} are presented.
Because all of the ISLAndS galaxies
have been observed to comparable depth, systematics in the models should have similar
impacts on all of the derived galaxy SFHs.  Thus, it is likely appropriate to make comparisons
using the upper panel.  Nonetheless, the lower panel shows the larger uncertainties
encountered when trying to account for systematics. Note that even with the larger uncertainties,
the six galaxies are shown to each have distinctive features in their SFHs.  The
effects of systematic uncertainties are discussed in more detail in the Appendix.

Presenting the SFHs as cumulative stellar mass fractions (as opposed to
the star formation rates - SFRs -  as a function of time) is the optimal way to compare
observations to theoretical models for several reasons.  Variations
in observed SFRs can be strongly affected by time binning and the changing time
resolution as a function of lookback time.  Often, it is possible to have very
different impressions of a single SFH simply by changing the time binning.
It is possible to match the observational time binning by reducing the resolution
in the models, but using the cumulative stellar mass fraction as the diagnostic
avoids this problem altogether.  It is also possible to compare galaxies at any
arbitrary value of the cumulative stellar mass fraction, as opposed to choosing
particular values to focus on.  In the comparisons that follow, we will use the
cumulative stellar mass fraction as the sole diagnostic.  

Note that there is one
obvious failing of the cumulative stellar mass fraction as the sole diagnostic, and
that is the lack of information about the absolute masses of the systems. In 
Figure \ref{f7}, we have added the stellar luminosities to the legend for ease
of comparison.  It is immediately clear that the mean age of the stellar populations
does not correlate with the present stellar luminosity.  
And~III, And~XV, and And~XXVIII have similar SFHs, despite significant 
differences in luminosity and distance from M31.
A more detailed comparison
of the SFHs as a function distance and luminosity is given in section \ref{seccomp}.

Figure \ref{f7} shows a variety of SFHs.
%While the comparison in the upper panel of Figure \ref{f7}, with only statistical 
%uncertainties, shows very clear differences between the SFHs, even the lower panel,
%including systematic uncertainties, shows that the SFHs are distinct. 
However, all six SFHs have the common properties of starting their star formation 
early (there is significant star formation in the earliest time bin for all
six galaxies), having at least 50\% of their stars formed by a lookback time of 9 Gyr,
and then ceasing all star formation by a lookback time of $\sim$ 6 Gyr. 
And~XV and And~XXVIII show the oldest mean stellar populations of the sample, but 
both are clearly
not consistent with a single age population and are clearly not consistent with
forming all of their stars before the epoch of reionization (as indicated in
Figure \ref{f7} by the shaded region).
As demonstrated in \citet{weisz2014a} and \citet{monelli2016}, And~XVI has a very extended SFH,
and the comparison in Figure \ref{f7} shows that it has the most extended star
formation in the sample.  
However, conspicuously absent from Figure \ref{f7} are galaxies with very 
late quenching times like the MW dSphs Carina, Fornax, and Leo~I.
A preliminary
comparison of the ISLAndS dSph SFHs with the MW dSph SFHs is given in section 
\ref{seccomp}.

\section{Possible Evidence for Synchronized Quenching}\label{secquench}

\subsection{The Quenching Time}

The definition of the quenching time for a galaxy is somewhat vague.
Part of this is operational, due to the presence of blue straggler stars.
These appear as a population aged $\sim$ 2 - 3 Gyr, amounting to as much as 
2 - 3 \% of the stellar mass; however, there is strong evidence in favor
of their interpretation as a population of merged primordial binaries with
the same age as the bulk of the old stellar population.   Therefore
defining the quenching time when the inferred SFR goes to zero can be 
misleading \citep[see discussion in][]{weisz2014c}.  Thus, typically a quenching 
time is set to a time when the cumulative star formation is approaching complete,
but something less than 95\% of the total mass of stars has formed.   For the purpose
of identifying an ``event'' that caused the quenching, it might make sense
to allow for star formation which follows the quenching.  For example,
\citet{ricotti2005} and \citet{bovill2011} promote the use of the age of 70\% of stars 
formed as a criterion for distinguishing true fossil of reionization from
other galaxies.  This 70\% criterion fails as a quenching identification
in galaxies with extended star formation.  For our purposes of searching 
for synchronized quenching, we will adopt an age when 90\% of the stars
have formed as a measure of the quenching time ($\tau_q$).  The 90\% serves as a very
robust measurement of the quenching time as it is not affected by the
presence of blue straggler stars.

\subsection{Synchronized Quenching?}

\citet{weisz2014a} noted that the SFHs of And~II and And~XVI showed nearly identical
quenching at a lookback time of approximately six Gyr.  \citet{weisz2014b} additionally 
noted that shallow CMDs of And~I, And~III, And~XI, and And~XII also indicated star formation
extending to intermediate ages, but the shallow depths of
these CMDs precluded strong inferences.  This possibility became even more 
intriguing with the addition of SFHs based on deep HST CMDs of NGC~147 and NGC~185
by \citet{geha2015}.  The SFH of NGC~147 is consistent with a quenching time
of six Gyr.  Although the SFH of NGC~185 shows it to be almost entirely old,
the uncertainties allow for continuing star formation up until a lookback time of
$\sim$ seven Gyr. However, as noted in the introduction, because the observed fields
in NGC~147 and NGC~185 are outside of the half-light radius, the interpretation and
comparison of their SFHs is a bit problematic.  

The addition of the new deep CMDs allow us to test for the possibility of 
synchronized quenching in the M31 satellites.  Indeed, the new SFHs give
no support for synchronized quenching.  
Specifically, 
And~III appears to have formed 90\% of its stars much earlier at $\sim$9 Gyr,
And~XV and And~XXVIII at a lookback time of $\sim$8 Gyr, and And~I at a lookback
time of $\sim$7.5 Gyr.

\citet{weisz2014a} offered two other explanations for the synchronized quenching
in And~II and And~XVI: (1) that both galaxies may have similar halo masses with similar
gas retention time scales \citep[implying significant differences in stellar mass at a 
fixed halo mass, see, e.g.,][]{boylan2011, garrison2014}, 
or (2) coincidental near passes with M31 
transforming previously gas-rich dwarfs into gas-poor dSphs via ``tidal stirring'' 
\citep{mayer2001a}.  Proper motions of these galaxies will be required to define their
orbits in order to distinguish between these options.

\subsection{Quenching by Reionization?}

Our understanding of the effect of reionization on dwarf galaxies is primarily driven 
by studies of MW companion galaxies with photometry that reaches below the oMSTO. 
Thus, it is important to know if the MW satellites are representative in this context 
or if the assembly of the MW and its satellite system is affecting our interpretations.
Our new imaging to below the oMSTO of a sample of M31 dwarfs allows us
to explore this question for the first time.  
The full details of the earliest SFHs will be deferred to a later study where
the inherent time resolution will be carefully modeled 
\citep[e.g.,][]{monelli2010b, monelli2010c, hidalgo2011, hidalgo2013},
and comparisons will be made to the most recent galaxy evolution models.
Nonetheless, we present here a brief
overview of the potential impact that these observations will have on the
study of the effects of reionization on the SFHs of dwarf galaxies.

\citet{weisz2014c} presented a review of the theoretical predictions and observational
tests connecting the cosmic UV background from reionization with the quenching of
star formation in dwarf galaxies.  We provide a very brief summary here. 
Quenching of star formation in dSphs by the cosmic UV background has been investigated 
multiple times \citep[e.g.,][]{efstathiou1992, bullock2000, benson2002, somerville2002,
benson2003, kravtsov2004, okamotot2008}.  With time, the expected imprint of the cosmic UV background
on the SFHs of dwarf galaxies has evolved.  Many early studies assumed that 
sufficiently low-mass galaxies would have their star formation  
permanently quenched by reionization on a very short time-scale.  Thus, it was tempting
to connect the lack of current star formation in the dSph galaxies 
with complete quenching by reionization.
\citet{grebel2004} pointed out that the observed SFHs of the Local Group dwarfs
did not support this simple picture.

Later models allowed for some star formation after reionization 
\citep[e.g.,][]{ricotti2005, gnedin2006, bovill2009, sawala2010} which reduced the
tension somewhat between the models and the observations of Local Group 
dSph galaxies.  However, recent, high resolution simulations of dwarf galaxies 
show that the impact of 
reionization can be much more subtle than originally imagined.
Supporting the original suggestion by \citet{susa2004} and \citet{dijkstra2004} 
that radiative transfer effects are important,  
recent work indicates that the cosmic UV background may suppress infall of fresh gas,
but is not 
%Specifically, a sophisticated treatment of the radiative transfer indicates that 
%the cosmic UV background may suppress infall of fresh gas, but is not
likely to boil away cold gas already present \citep[e.g.,][]{onorbe2015}.
This means 
that searching for the signature of reionization in the SFHs of dwarf galaxies
is much more complicated than originally envisioned. The simulations of \citet{wheeler2015}
provide a dividing line of the peak virial mass, M$_{peak}$, $=$ 5 $\times$ 10$^9$ M$_{\odot}$, below which
their model galaxies only show star formation to be entirely quenched by z $\sim$2 (a
look back time of $\sim$ 10 Gyr).  

In their models, M$_{peak}$ $=$ 5 $\times$ 10$^9$ M$_{\odot}$
corresponds to present day stellar masses of $\sim$  3 $\times$ 10$^5$ M$_{\odot}$.
In this regard, assuming a V-band mass-to-light ratio of 1, And~VI is below this dividing line,
And~III, And~XV, and And~XXVIII are close to this dividing line, and And~I and And~II are 
well above this division.

It is clear that the SFHs of all six galaxies are extended and inconsistent 
with a single age stellar population.  How does this compare with expectations? 
In the models of \citet{wheeler2015}, M$_{peak}$ $=$ 5 $\times$ 10$^9$ M$_{\odot}$
corresponds to present day stellar masses of $\sim$  3 $\times$ 10$^5$ M$_{\odot}$.
In this regard, assuming a V-band mass-to-light ratio of 1, And~VI is below this dividing line,
And~III, And~XV, and And~XXVIII are close to this dividing line, and And~I and And~II are
well above this division.
Table~1 presents the circular velocities measured at the half-light radii for the
six ISLAndS galaxies as reported in \citet{collins2014} and \citet{tollerud2014}.  These
range from $\sim$ 6 km~s$^{-1}$ (And~XV) to $\sim$ 18 km~s$^{-1}$ (And~I).
The values of M$_{tot,1/2}$ presented in Table~3 from the \citet{walker2009} mass estimator 
\citep[which is consistent within the observational uncertainties with that of][]{wolf2010}
range from $\sim$3 $\times$ 10$^6$ M$_{\odot}$ (And~XV and And~XVI) 
to $\sim$3 $\times$ 10$^{7}$ M$_{\odot}$ (And~I and And~II).
Thus, from stellar velocity dispersions, And~XV and And~XVI are right at the limit of where galaxies are expected
to have their star formation quenched by reionization.  The SFH of And~XV is
possibly marginally consistent with the models of \citet{wheeler2015}, showing the star 
formation to be almost entirely quenched at a lookback time of 8 Gyr.  However, And~XVI shows 
the most extended SFH of the sample. As pointed out by \citet{weisz2014a} and 
\citet{monelli2016}, the 
SFH of And~XVI, with its large intermediate age population and continuous star formation, 
is inconsistent with quenching by reionization.  The lack of correlation of quenching time
with measures of the halo mass (e.g., the values of V$_{c,1/2}$ listed in Table~1), put
strong constraints on the potential role of quenching by reionization in this 
galaxy sample. 

Much of the literature discussion focuses on reionization quenching, which completely
terminates star formation at the reionization epoch or shortly thereafter.  However,
we would like to also look for possibly less catastrophic impacts of reionization, where
star formation is temporarily slowed or paused and resumes at a later time.
One thing that is interesting about the SFH of And~XVI is that the SFR decreases
immediately after the post-reionization period.  We may be seeing evidence of
the impact of the ultraviolet background on the star formation in And~XVI, but 
the result is not a complete quenching. 
Several
%It is interesting that the SFH of And~XXVIII appears a little bit different from the others
%in that its SFR increases in the immediate post-reionization period, which is the
%opposite of the other galaxies.  
of the other SFHs in Figure~\ref{f7} show an 
inflection in their cumulative stellar mass fractions indicating a slowdown 
in the formation of stars immediately post-reionization.  
This is certainly interesting in the context of any kind of
link between reionization and suppression of SFR.
It is also interesting that the SFH of And~III appears a bit different from the other galaxies
in that its SFR increases in the immediate post-reionization period
(and demonstrates that the slowdown seen in the other galaxies is
not a systematic effect from the CMD modeling).  
We will focus on these interesting 
features in our detailed studies of the earliest SFHs, with special focus on the 
robust characteristics that are independent of the stellar evolution library.

\section{Is There a Thin Plane/Non-thin Plane Dichotomy?}\label{secplane}

\citet{ibata2013} discovered that about half of the satellites of M31 reside in a planar subgroup.
This vast, thin structure is at least 400 kpc in diameter, but less than 14 kpc
in width. The member satellites rotate about M31 in the same sense. The reality and probability
of this plane has been discussed at length, especially with regard to whether such a structure
is consistent with galaxy formation in a $\Lambda$CDM cosmology \citep[see, e.g.,][]
{hammer2013, ibata2014, bahl2014, pawlowski2014, gillet2015, buck2015, libeskind2015, phillips2015}.

\citet{collins2015} have investigated the possibility that in-plane galaxies and off-plane galaxies may have formed and evolved 
differently.  The different evolutionary paths could result in different chemistries, dynamics, and 
star formation histories. From a comparison of the relationships between the velocity dispersions, 
masses, half-light radii, luminosities, and metallicities available for all of the M31 dwarf spheroidal 
galaxies, \citet{collins2015} found the on- and off-plane Andromeda dwarf galaxies to be indistinguishable 
from one another.  

One characteristic that was not available to \citet{collins2015} was the lifetime
SFHs of the M31 satellites.  For the sample here, three of the galaxies (And~I, And~III, and And~XVI) 
are in the plane and three of the galaxies (And~II, And~XV, and And~XXVIII) are not
(see Figure \ref{f2}).  Comparing 
the star formation histories, we also find no evidence for differences between the on- and off-plane
galaxies.  Specifically, And~II and And~XVI are the two galaxies with the most extended SFHs, yet one
is off-plane and the other is in-plane.  And~III, And~XV, and And~XXVIII are the three galaxies with
the earliest truncation times, and  
one is in-plane and the other two are off-plane.  It would appear that the physical process leading to 
the distinction of in-plane and off-plane galaxies does not leave an imprint on any of the
intrinsic properties of the galaxies.

\section{Comparing The Lifetime SFHs of the M31 and Milky Way Satellites}\label{seccomp}

We reserve a detailed examination of the earliest SFHs of the ISLAndS galaxies for
a follow-up study.  The SFHs of the first few Gyr entails precise modeling 
of the intrinsic time resolution of the observations 
\citep[see discussions in][]{hidalgo2011, aparicio2016} and we are also planning to use 
the latest updates for the MESA \citep{dotter2016} and BASTI stellar evolution libraries.
Nonetheless, interesting comparisons of the very robust late-time features of the
SFHs can be made at this time with great confidence.

\citet{weisz2014a} provided an initial glimpse at comparing SFHs of M31 and MW dSphs by
comparing the SFHs of And~II and And~XVI to MW dSphs with comparable luminosities and
at comparable host distances.  As the number of galaxies with deep CMDs allowing for 
secure SFHs increases, it is desirable to produce a diagnostic diagram in which
the comparison is simplified.  
Two key characteristics of a SFH are
the time required to produce 50\% ($\tau_{50}$) of the total star formation and the 
quenching time ($\tau_{q}$).
The time for the production of 50\% of the total stars can be measured robustly and 
gives a first order characterization separating galaxies that form quickly from
those with a more steady build up of stars.  As discussed earlier, the time associated 
with 90\% of the total star formation gives a good approximation for the quenching time, 
which is free from contamination by the presence of blue straggler stars. 

We can use these two characteristics to construct three diagnostic diagrams to 
compare the MW and M31 dSph SFHs:
(1) a comparison of the times when 50\% and 90\% of the stars are formed as a 
function of current distance to the host galaxy;
(2) a comparison of the times when 50\% and 90\% of the stars are formed as a 
function of the luminosity of the galaxy; and
(3) a direct comparison of the times when 50\% and 90\% of the stars are formed.
Here we use the SFHs for the MW dSphs taken from \citet{weisz2014b} to compare to 
the ISLAndS dSphs studied here.  There are two advantages to this MW dSphs
comparison set. First, both studies have used the Padova stellar evolution library, 
so the systematic offsets resulting from choice of library, while not as important for 
the later times concerned here, are minimized.  Second, the SFHs are calculated in
an identical, uniform way, so that direct comparisons are valid. 
In order to provide a fair comparison, we restrict the MW comparison sample
to a range in an absolute luminosity of $-$14 $\le$ M$_V$ $\le$ $-$6.  This range 
encompasses the range from the ISLAndS sample.

One disadvantage of the \citet{weisz2014b} sample is the small HST fields of view (and 
correspondingly relatively small number of stars measured) for the closest MW companions.
In this regard, ground-based observations extending below the oMSTO 
\citep[e.g.,][]{okamoto2008, okamoto2012} can be competitive for providing accurate 
SFHs.  
As part of the ISLAndS project, we are re-analyzing a large collection of
ground-based observations of MW companion galaxies to produce a single,
uniform analysis.  For this reason, we consider the following comparisons preliminary,
and will re-visit the exercise with well-defined uncertainties for all MW companion galaxies 
at a later date.  Nonetheless, we feel that the patterns revealed in this study are
quite robust, even if the individual data points may experience small shifts in the final 
analysis.  Note also that these comparisons are all in the regime of small 
number statistics.  A single addition or subtraction of a galaxy can 
make a significant impact on the impression of these comparisons.  Thus, it
is best to focus on broad trends.

\begin{figure}[h]
\centering
\includegraphics[width=8.5cm]{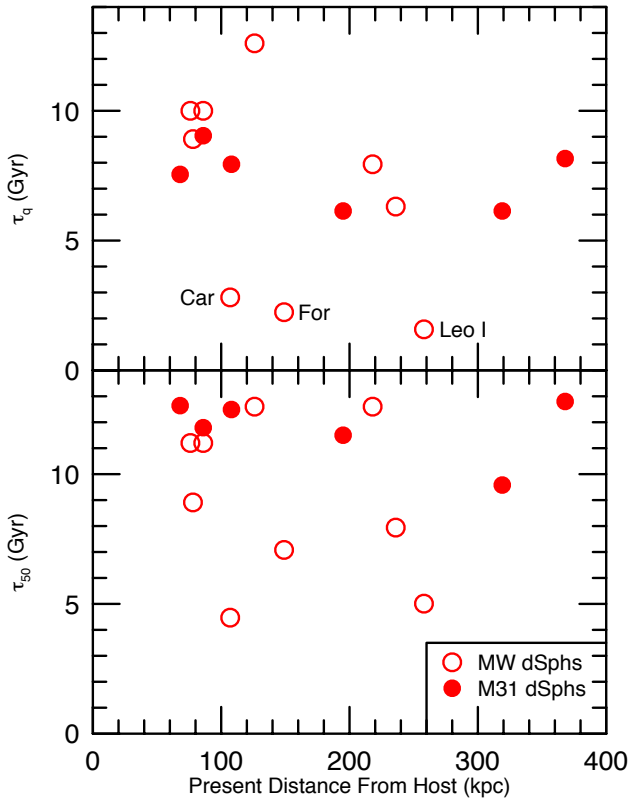}
\caption[ ]{A comparison of the lookback times for 50\% ($\tau_{50}$, lower panel) 
and 90\% ($\tau_{q}$, upper panel) of
the stars to form as a function of distance from the MW for the MW dSphs taken from
\citet{weisz2014b} and the ISLAndS dSphs studied here.
Note that there is only a
weak trend in the importance of intermediate age populations as a function of
present distance from the host galaxy for both samples. Note also that there are no
M31 dSphs with very late ($\le$ 5 Gyr) quenching times.  The three MW dSphs with
very late quenching times (Carina, Fornax, and Leo~I) are labeled.
}
\label{f8}
\end{figure}

\subsection{Trends with Host Distance}

Figure~\ref{f8} shows the comparison of the lookback times for 90\% ($\tau_{q}$; upper panel) and 
50\% ($\tau_{50}$; lower panel) of the stars to form 
as a function of distance from the host galaxy for the MW dSphs and
the ISLAndS dSphs studied here.  Starting with the top panel, 
note that in this representation, the trend for increasing
importance of an intermediate age population with distance is not very strong
for the MW dSphs.  The MW dSphs show a weak trend in $\tau_{q}$ with distance 
from host with very large scatter.  In comparison,  
the ISLAndS dSphs generally also show a weak trend for $\tau_{q}$ to decrease with
present host distance with less scatter.  

The impressive part of this comparison
is the complete lack of galaxies in the ISLAndS sample showing
a very late quenching time ($\le$ 5 Gyr).  
The late quenching MW satellites stand out as a separate
population in the upper panel of Figure~\ref{f8}. 
This is the most significant difference between the two populations.   
(Note the possible exception that, at a higher luminosity, NGC~147 does have 
a late quenching time \citep{geha2015}.)
The three MW dSphs galaxies with 
very late quenching times are Carina, Fornax, and Leo~I, and these are labeled
in Figure~\ref{f8}.    

In the lower panel of Figure~\ref{f8}, showing $\tau_{50}$ as a function of distance
from host, the MW dSphs show no trend with distance at all.  The ISLAndS sample 
shows a mild trend for decreasing $\tau_{50}$ with increasing present distance.
All of the galaxies in the ISLAndS sample have formed 50\% of 
their stars before a lookback time of 9 Gyr.

The trend for the stellar content of faint Local Group dwarfs to correlate with distance 
from the Galaxy identified by \citet{vdbergh1994b}, is not very apparent when restricted
to the dSph population in the present luminosity range.  The interpretation of this
trend is complicated by the fact that these galaxies can be on non-circular orbits, and
their present host distances may not be representative of their locations at the times
that we are comparing them.  As pointed out by \citet{bullock2005} the average infall
times for surviving satellite systems have a median $\sim$5 Gyr in the past, so the
typical surviving satellite would have been outside of the gravitational influence of
the host galaxy when the bulk of its star formation occurred.  Detailed modeling by
\citet{wetzel2015a} has confirmed this picture and further highlighted the 
``pre-processing'' by other gravitational interactions before falling into the 
present day host.  \citet{benitez2013} have also pointed to the potential processing 
of dwarf galaxies by ``cosmic web stripping,'' which adds further complications in
understanding the histories of dwarfs galaxies.  The broad trend observed by
\citet{vdbergh1994b} comparing diverse morphological types should not necessarily 
be expected to be obvious in a sample restricted to only dSph galaxies with present
day distances.  Clearly, adding orbital information is highly desirable to 
allow better formed questions to be investigated.

\begin{figure}[h]
\centering
\includegraphics[width=8.5cm]{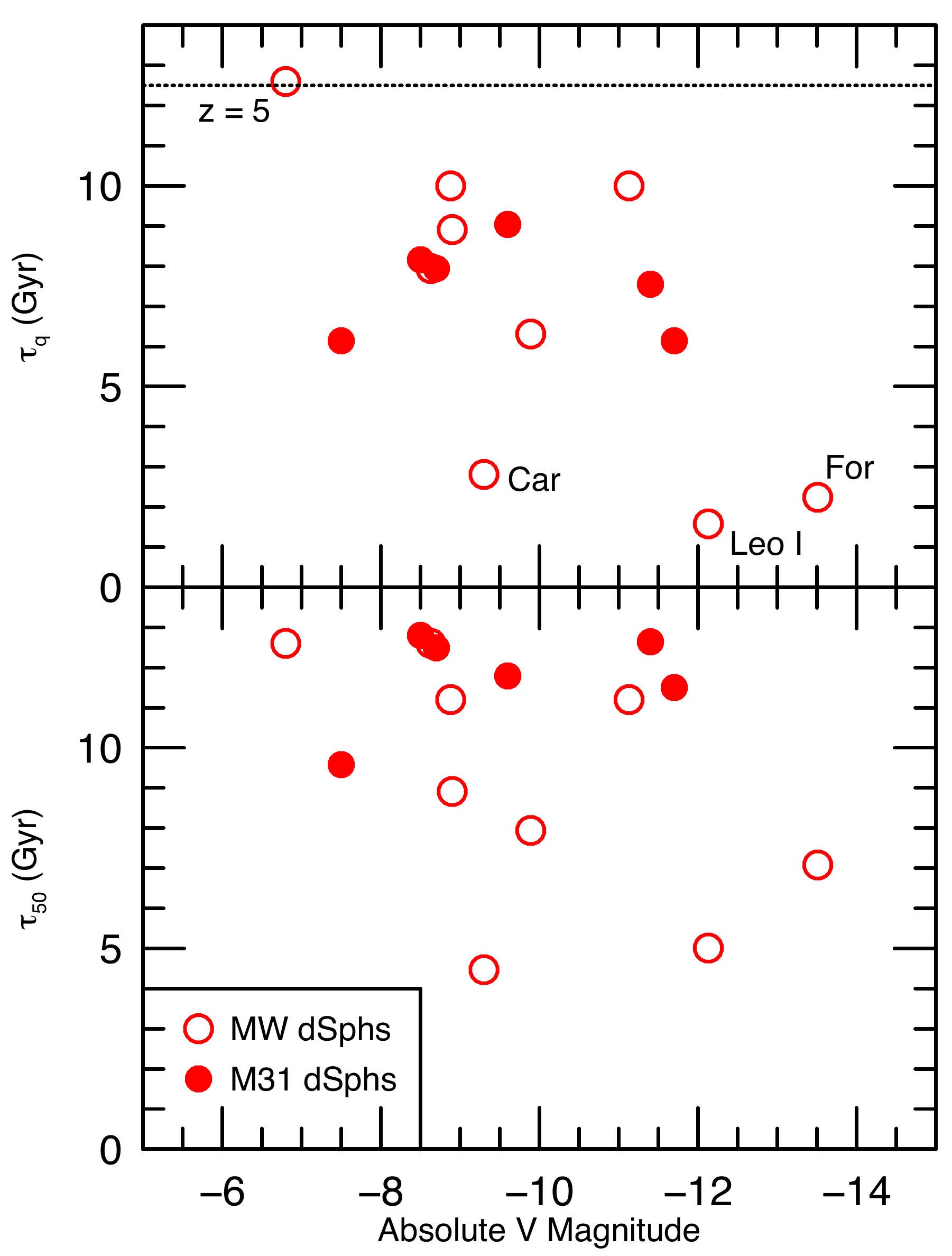}
\caption[ ]{ A comparison of the lookback times for 90\% ($\tau_{q}$; upper panel) and
50\% ($\tau_{50}$; lower panel) of the stars to form
as a function of galaxy luminosity for the MW dSphs taken from
\citet{weisz2014b} and the ISLAndS dSphs studied here.
The dotted line in the upper
panel corresponds to a redshift of 5 when reionization has been fully completed \citep{becker2015}.
Galaxies which were completely quenched by reionization would lie above this line.
Note that there is a
trend for later quenching with increasing luminosity for both dSph populations.
Note also the difference between the MW and M31 populations in their lookback times for
50\% of the the stars to form.
}
\label{f9}
\end{figure}

\subsection{Trends with Galaxy Luminosity}

It is possible that Figure~\ref{f8} could be biased by not comparing similar 
luminosity MW dSphs with M31 dSphs.  Although the range in absolute luminosity was
restricted, that does not necessarily mean there is no bias in comparing the
two samples. 
To test for such a possible sample bias, in Figure~\ref{f9} we plot the characteristic times
versus the galaxy V-band luminosities.  The comparison for quenching times is shown in
the top panel. In the top panel we have added a dotted line 
corresponding to a redshift of 5 when reionization has been fully completed \citep{becker2015}.
Galaxies which were completely quenched by reionization would lie above this line.
Only the very lowest luminosity galaxy in the MW sample (Hercules) is consistent with
complete quenching by reionization.

Overall, the top panel of Figure~\ref{f9} shows a stronger trend in the MW dSphs with luminosity than
we saw when plotted as a function of present distance.  The more luminous MW dSphs tend, 
on average, to have later 
quenching times.  The baseline in luminosity for the ISLAndS sample is slightly smaller, 
but there is little evidence of a trend of quenching time with
luminosity in the ISLAndS sample.  Indeed, the M31 dSphs appear to be a much more homogeneous
set in this regard. 
From Figure~\ref{f9}, it is clear 
that the lack of very late quenching times in the ISLAndS sample is not due to the
luminosities of the sample galaxies. 

The bottom panel of Figure~\ref{f9} essentially reflects the trends seen in the 
top panel.  The $\tau_{50}$ values for the MW dSphs show a trend with luminosity in 
the sense of later star formation for higher luminosities, while the ISLAndS sample
gives no evidence for a trend.  How much of the two trends for the MW dSphs in 
Figure~\ref{f9} is due to the presence of the three late quenching galaxies is
difficult to say. If they are removed from the comparison, then the five 
remaining galaxies still show a trend, albeit very much weaker, where the 
impression of a trend can be altered by the presence/absence of a single 
galaxy.  In this regard, the absence of a trend in luminosity in the ISLAndS 
sample cannot be definitively see as different from the MW sample.  The main
difference remains that the ISLAndS sample has no very-late quenching 
galaxies, while these are common in the MW satellites of the same luminosity.

Comparing $\tau_{q}$ values (top panel) with the $\tau_{50}$ values (lower panel), 
there is another potentially interesting difference between the two samples.
The M31 satellites have, on average, produced 50\% of their stars much faster 
that the MW satellite. On the other hand, the M31 satellites have quenching times  
comparable to those of the MW satellites. 
Taken at face value, 
then this could mean that the M31 satellites 
experienced a much stronger early impact than the MW dwarfs.  That is,
star formation was greatly suppressed in the M31 satellites and then only
returned back to comparable star formation rates after a significant delay
(as seen in Figure~\ref{f7}).  This difference could be due to
the delayed effects of reionization, i.e., although neither sample is showing 
complete quenching by reionization, perhaps the two have been impacted 
differently by reionization.  Alternatively, this differnce could be reflecting the 
impact of the parent galaxies on the early evolution of the satellites.  With
such small samples and recognizing the complications of orbital histories, it
will be difficult to draw definitive conclusions.  Nonetheless, there 
appears to be evidence that the evolution of the M31 dSphs was more 
uniform than the evolution of the MW dSphs.

\begin{figure}[h]
\centering
\includegraphics[width=8.5cm]{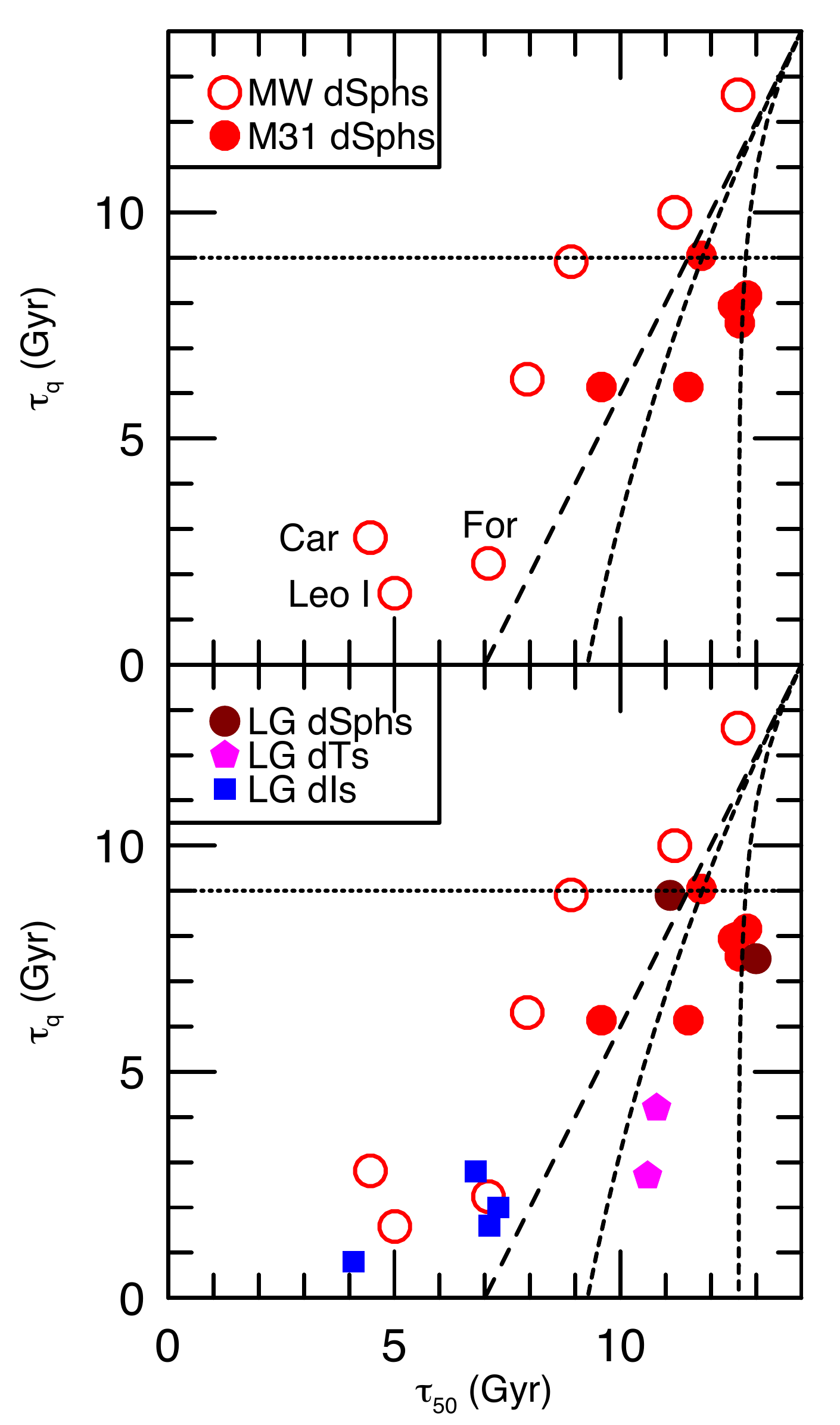}
\caption[ ]{Upper panel: A comparison of the lookback times for 90\% ($\tau_{q}$) 
and 50\% ($\tau_{50}$) of the stars to form for the MW dSphs taken from 
\citet{weisz2014b} and the ISLAndS dSphs studied here.  The long dashed line represents the
relationship for constant star formation ended by instantaneous quenching.  The two
short dashed lines are for exponentially decreasing star formation ($\tau$ models) ended by instantaneous 
quenching (with time scales of 2 Gyr and 10 Gyr).  The dotted line approximates the 
division between ``fast dwarfs'' and ``slow dwarfs'' as defined by \citet{gallart2015}.
The ensemble of MW satellite galaxies is roughly consistent with constant star formation with
rapid quenching.  The M31 satellite galaxies are roughly consistent with $\tau$ models.
The M31 satellite galaxy sample shows no late quenching galaxies as observed in the
MW sample.
Lower panel: We add to the top panel values of $\tau_{q}$ and $\tau_{50}$ for
isolated Local Group dwarfs with oMSTO photometry.  The isolated dSphs 
Cetus and Tucana fit in with the satellite dSphs, the dI galaxies overlap with
the late quenching MW dSphs, and the isolated dT galaxies define their own region.
} 
\label{f10}
\end{figure}

\subsection{Comparing SFH Characteristics}

In this last section, we compare the values of $\tau_{q}$ versus $\tau_{50}$.
In the top panel of Figure~\ref{f10}, we plot $\tau_{q}$ versus $\tau_{50}$ for 
the MW and M31 satellite samples.  
Here we see the very
interesting trend that $\tau_{q}$ and $\tau_{50}$ are relatively well correlated for
both the MW and the M31 dSphs.  
This is not totally unexpected, as $\tau_{q}$ must always be greater than or equal to $\tau_{50}$,
but galaxies which form a substantial fraction of their stars early and then quench 
very late (i.e., galaxies which would be in the lower right hand corner of
figure~\ref{f10}) are, in principle, possible, but not observed.
The general trend is, to first order, consistent with relatively constant star formation
ended by a quenching event.  The long dashed line in Figure~\ref{f10} shows $\tau_{50}$ $=$ $\tau_{q}$/2, 
which would be the result of constant star formation followed by rapid quenching.
There is good evidence for a rapid transformation of star forming satellite dwarf galaxies 
into dSphs after falling into the host galaxy halo \citep{wetzel2015b}, so the 
assumption of an instantaneous quenching is supported.  For the late-quenching galaxies,
quenching by reionization is certainly not a factor, and environmental quenching
associated with infall seems most likely.  Of all of the possible
ways to remove the gas, ram pressure stripping is the fastest \citep{mayer2007}.

Star formation in galaxies is often modeled as exponentially decreasing ($\tau$ models), so it is 
of interest to see where these models would lie in this diagnostic diagram.  The 
dotted lines show tau models with time constants of 2 and 10 Gyr.  Since increasing the
time constant will only more closely approximate constant star formation, it is 
clear that constant star formation is a better approximation to the SFHs of  
the MW dSphs.  On the other hand, most of the M31 satellites can be found lying
between the two $\tau$ models.  \citet{weisz2014b} found that the SFHs of dSph galaxies could
be approximated by exponentially declining star formation with a tau of 5 Gyr.
It appears from Figure~\ref{f10} that constant star formation is a better characterization
for the MW satellites, but that the M31 satellites might be better described with
a range of $\tau$ models.

\citet{gallart2015} have proposed that the dSphs can be divided into ``slow'' and ``fast''
classes based on their SFHs. They propose that the distinction between a fast and slow dwarf 
galaxy primarily reflects the characteristic density of the environment where they form.
While there is no quantitative definition of ``fast''
and ``slow,'' all of the example ``fast'' galaxies are quenched by a lookback time of 9 Gyr. 
We have added a dotted line to Figure~\ref{f10} to indicate this division.
Figure~\ref{f10} shows that while there are MW companions above this line, 
most of the M31 dSph satellites lie below this line.
There is no evidence of a clear dichotomy in Figure~\ref{f10}, and most of the 
M31 dSph satellites in our sample lack the extremes of a short dominant episode of 
star formation or a very extended star formation history.  However, the connection 
of fast star formation to evolution in denser environments is not
ruled out by Figure~\ref{f10}, only the case for a clean separation.

The top panel of Figure~\ref{f10} reinforces the main conclusion from this section that 
the M31 and MW dSph
populations distinguish themselves by the complete absence of late quenching dSphs in
the M31 population.  Of course, the ISLAndS sample is only representative and not
complete, but there has been no selection bias which could have led to this
result.  The M31 and MW dSph samples overlap in distance from host and galaxy luminosity,
so there must be some other parameter of importance.  A secondary, and more subtle
difference is the segregation of the M31 and MW dSph populations in this diagram.
To first order, the M31 dSphs lie on or below the long dashed line indicating 
constant star formation, while the MW dSphs are all above this line.  This is a
re-manifestation of the trend already noted that the M31 dSphs produced half of their
stars earlier than the MW dSphs and then were quenched at approximately the
same time (discounting the late-quenching MW dSphs). 

In the bottom panel of Figure~\ref{f10}, we have added the results from the 
SFHs from the six isolated Local Group dwarfs from the LCID study 
\citep[from Figure 10 in][]{skillman2014} consisting
of two dSphs (Cetus and Tucana), two dTs (LGS-3 and Phoenix), and two dIs
(Leo A and IC 1613).  We have also added two other isolated dIs with comparable 
deep HST photometry: Leo T \citep{weisz2012} and Aquarius \citep{cole2014}.

The Local Group isolated dSphs Cetus and Tucana fit in with the distribution of companion dSphs.
From the point of view of SFHs, there seems to be little to distinguish the 
isolated dSphs from companion dSphs.  Since the radial velocities of Cetus and
Tucana \citep{lewis2007, fraternali2009} are consistent with orbiting within
the virial volume of the MW \citep{teyssier2012}, it supports the premise
that there is a commonality to evolutionary paths of all of the LG dSphs.   

The Local Group dwarf transition galaxies (dTrans, or dT) galaxies are defined 
by the presence of recent star formation but
the lack of very recent massive star formation (i.e., the lack of detected \ion{H}{2} regions).
The nature of dT galaxies has been discussed by \citet{skillman2003} who found that
most dTs are consistent with normal dI galaxies exhibiting temporarily interrupted star formation,
but that their observed density-morphology relationship indicates that environmental processes 
may play a role in causing their lower present day SFRs.
From SFHs, \citet{weisz2011} found the majority of the dTs to be low-mass dIs that simply 
lack H$\alpha$ emission, but that some dTs (like Phoenix) have remarkably low gas fractions.
The LCID project provided very detailed SFHs for Phoenix \citep{hidalgo2009} 
and LGS-3 \citep{hidalgo2011} and they are included
in the bottom panel of Figure~\ref{f10}.  Interestingly, the isolated dT galaxies define their 
own region in the diagnostic diagram in Figure~\ref{f10}. 
Although there are no dSphs in the early build-up and late quenching corner of
the diagram, nature is able to produce galaxies in this parameter space through the
process or processes that produce dT galaxies.   
The dT galaxies are characterized by significant early star formation which has 
decreased in amplitude in time, and thus, they occupy
a region of the diagram that fits well with $\tau$ models.   

Finally, the dI galaxies overlap with the late quenching MW dSphs  
\citep[Note that we have included Leo~T with the
dI galaxies with the justification that Leo~T is simply a low-mass dI
lacking H$\alpha$ emission.  In many regards, Leo~T is similar to the dI Leo~P
with the exception that Leo~P does have an \ion{H}{2} region,][]{mcquinn2015}.
The overlap of the dIs with the late quenching dSphs is really quite remarkable
and suggests that just a few Gyr ago the late quenching MW dSphs would 
have been identical to today's dIs.  This reinforces the idea of an
environmental evolutionary link between the two and the conclusion 
that the current morphological type is not necessarily a good diagnostic of the
intrinsic, long term evolution of a galaxy properties \citep[e.g.,][]{gallart2015}.  
The measurement of Leo~I's proper motion 
by \citet{sohn2013} shows a coincidence of the quenching time with the 
time of the orbital perigalacticon, which suggests that Leo~I may offer a 
prototype for the conversion of star forming to gas-free dwarf galaxies 
due to processes associated with infall into the host galaxy's virial
volume.
\citet{wetzel2015a} have modeled the infall histories of dwarf galaxies and 
identified two sources of strong environmental influence, the virial volume
of the host and ``group pre-processing.''  The challenge is now to combine
SFHs and reconstructed orbits to determine whether these models provide 
an adequate description of the galaxies in the Local Group.  Additional
modeling that can reproduce satellite populations with and without late
quenching satellites will be of extreme interest.  
  
\section{SUMMARY AND CONCLUSIONS}\label{seccon}

We have presented HST ACS and WFC3 observations of six M31 dSph galaxies as part of
the Initial Star formation and Lifetimes of Andromeda Satellites (ISLAndS) project.
These six galaxies comprise a representative sample of Andromeda dSph satellite 
companion galaxies.  The main goal of this program is to determine whether the star 
formation histories (SFHs) of the Andromeda dSph satellites are significantly different 
from those of the Milky Way.

Our observations reach the oldest main sequence turn-offs, allowing a time
resolution at the oldest ages of $\sim 1$ Gyr.  We find that the six dSphs present a
variety of SFHs which are not strictly correlated with luminosity or present distance
from M31. Specifically, we find a broad range in quenching times, but all quenching
time are earlier than $\sim$6 Gyr ago.  
In agreement with
observations of Milky Way companions of similar mass, there is no evidence of complete
quenching of star formation by the cosmic UV background responsible for reionization, but
the possibility of a degree of quenching at reionization cannot be ruled out. 

We do not find
significant differences between the SFHs of the three members of the vast, thin plane of
satellites discovered by \citet{ibata2013} and the three off-plane dSphs in our
sample.  This is
further confirmation of the conclusion by \citet{collins2015} that there is no
evidence of physical differences between the in-plane and out-of-plane galaxies. 

For a simplified comparison of SFHs and a preliminary comparison between the MW and M31
dSphs, we concentrate on the lookback times when 50\% and 90\% of the total stars are formed.
A preliminary comparison of the SFHs of the ISLAndS dSphs with those Milky Way 
dSph companions with similar luminosities and host distances 
shows that there is one primary significant difference using this diagnostic.
There are no very late quenching ($\tau_{q}$ $\le$ 3 Gyr)  dSphs in the ISLAndS sample.  
Thus, modeling that can reproduce satellite populations with and without late
quenching satellites will be of extreme interest.
A secondary and more subtle
difference is the segregation of the M31 and MW dSph populations in this diagram.
The M31 dSphs produced half of their stars earlier than the MW dSphs and then were 
quenched at approximately the same time (discounting the late-quenching MW dSphs).
Interestingly, to first order, the ensemble of M31 dSphs are consistent with 
$\tau$ models while the MW satellites are best described as having nearly 
constant star formations rates until being rapidly quenched.

\acknowledgments

Support for this work was provided by NASA through grants GO-13028 and GO-13739
from the Space Telescope Science Institute, which is operated by
AURA, Inc., under NASA contract NAS5-26555.
Support for DRW is provided by NASA through Hubble Fellowship grant
HST-HF-51331.01 awarded by the Space Telescope Science Institute.
The computer network at IAC operated under the Condor software license has been used.
Authors SH and AA are funded by the IAC (grant P3/94) and, with SC, by the Economy 
and Competitiveness Ministry of the Kingdom of Spain  (grant AYA2013-42781P). 
%Authors SH and AA are funded by the IAC (grant P3/94) and, with SC, PS, and EB,
%by the Science 
%and Technology Ministry of the Kingdom of Spain (grant AYA2007-3E3507) and
%Economy and Competitiveness Ministry of the
%Kingdom of Spain (grant AYA2010-16717).
Authors CG, CEM-V, and MM are funded by the IAC grant P/301204 and by the Spanish Ministry of
Economy and Competitiveness (MINECO) under the grant (project reference AYA2014-56795-P).
%Authors CG and MM are funded by the IAC grant P/301204 and by the Economy and 
%Competitiveness Ministry of the Kingdom of Spain (grant AYA2014-56795). 
This research has made use of NASA's Astrophysics Data System
Bibliographic Services and the NASA/IPAC Extragalactic Database
(NED), which is operated by the Jet Propulsion Laboratory, California
Institute of Technology, under contract with the National Aeronautics
and Space Administration.

\appendix

\begin{figure}[h]
\plottwo{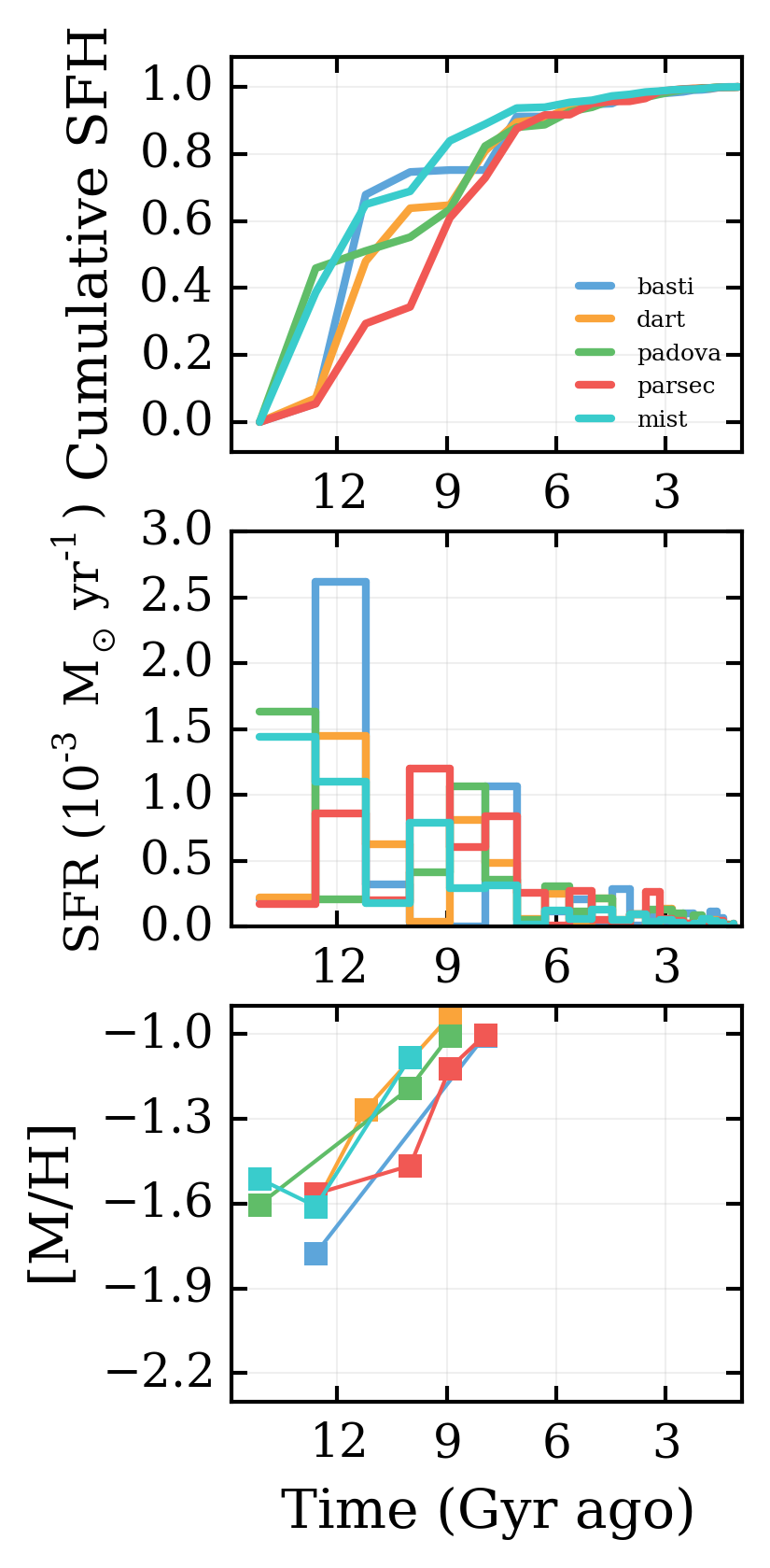}{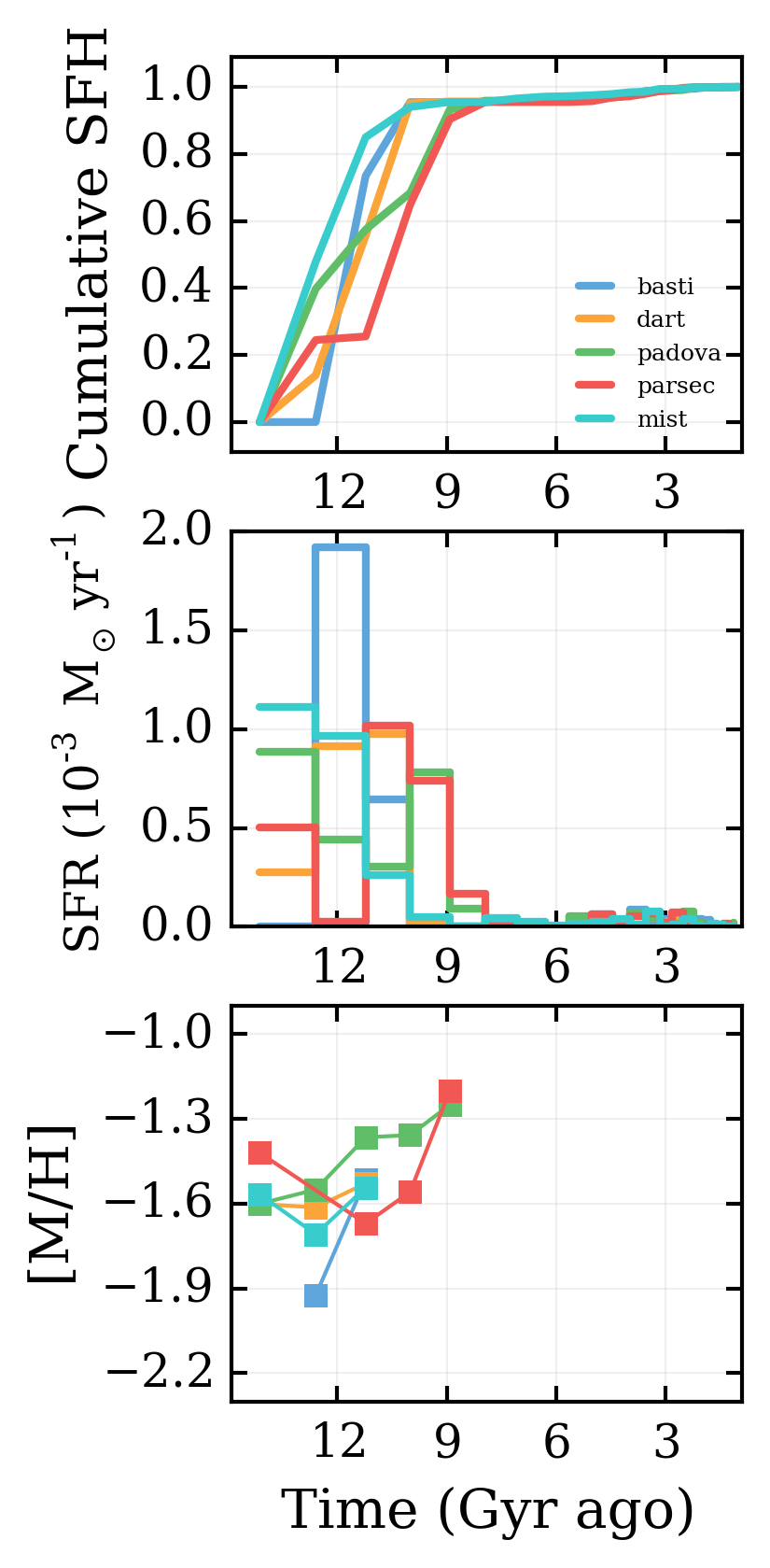}
\caption[ ]{ The cumulative SFHs, SFRs, and AMRs
for And II (left panel) and And III (right panel) derived using five different 
stellar evolution libraries \citep[PARSEC, Padova, Dartmouth, BASTI, and MIST:][]{bressan2012, girardi2010,  
dotter2008, pie_etal2004, choi2016}.  While all five solutions are similar, there are systematic differences
which are at least as large as the statistical uncertainties in the solutions.  Note that the differences 
are consistent between the galaxies, i.e., the MIST and BASTI solutions form stars the fastest, the Padova
and Dartmouth solutions are intermediate and the PARSEC solutions form the stars the slowest.  Thus, comparisons
conducted using a single stellar evolution library are likely to result in robust differences/similarities,
but the absolute time frame will depend on the choice of stellar evolution library.
} 
\label{fa1}
\end{figure}

\section{The Impact of Different Stellar Libraries on SFHs}\label{appen} 

There are three main sources of uncertainties
in deriving SFHs \citep[see, e.g., discussion in][]{apa_hid2009}:
the input observational uncertainties, the statistical
uncertainties of the solution, and the systematic uncertainties of the adopted isochrones from 
the stellar evolution libraries.
In a specific test of creating synthetic photometry from one stellar evolution library and
using a different stellar evolution library to derive a SFH, \citet{apa_hid2009} showed that
the systematic uncertainties associated with the adopted isochrones from
the stellar evolution libraries dominated
those of the numerical methodology or the observation uncertainties.
They recommend always using more than one stellar evolution library when analyzing
a real population to test for these effects.
\citet{weisz2011} and \citet{dolphin2012} have also shown that, for sufficiently
deep photometry, the stellar evolution uncertainties, as approximated by
differences between stellar evolution model and their resulting libraries,
represent the dominant systematic uncertainty for derived SFHs.
\citep[See also the discussion in][concerning the effects of varying heavy element
abundance patterns.]{dotter2007}

In order to give a sense of the importance of the choice of the stellar evolution library,
here we present comparisons for two of the ISLAndS sample (And~II, one of the two galaxies
with extended star formation and a quenching time of $\sim$6 Gyr, and And III, 
the earliest quenched galaxies of the sample, with a quenching time of $\sim$9 Gyr). 
Figure~\ref{fa1} shows the cumulative SFHs, SFRs, and AMRs
for And II (left panel) and And III (right panel) derived using five different
stellar evolution libraries 
\citep[PARSEC, Padova, Dartmouth, BASTI, and MIST:][]{bressan2012, girardi2010,
dotter2008, pie_etal2004, choi2016}.  While all five solutions are similar, there are systematic differences
which are at least as large as the statistical uncertainties in the solutions
(compare with Figure~\ref{f7}).  
These differences come from at least three different sources.  First, the 
stellar libraries can change as a result of updating input data such as 
nuclear reaction rates.  Secondly, the codes can differ in how to manage
the efficiency of non-canonical processes such as core overshooting.
Thirdly, different libraries make different 
assumptions about certain model specifications
such as the adopted heavy element solar mixture, the initial He abundance at the lower
metallicity (the primordial helium abundance), the assumed helium-enrichment ratio dY/dZ.
\citet{choi2016} present a useful comparison of some of the differences between
existing databases.

Dramatic differences in the bin-to-bin SFRs are minimized in the 
cumulative SFHs plots, but the positions of steep changes in the SFHs can
be seen to shift systematically with choice of stellar evolution library.
Note that the differences
are consistent between the galaxies, i.e., the MIST and BASTI solutions form stars the fastest, 
the Padova and Dartmouth solutions are intermediate and the PARSEC solutions form the stars 
the slowest.  
Thus, when comparing SFHs between galaxies or between groups of galaxies, 
it is imperative to approach the data with as uniform a methodology as 
possible, and especially, to use the same stellar libraries.  Ideally,
comparisons should be conducted using a variety of libraries such that 
a sense of the systematic uncertainty associated with the choice of
the library can be achieved.  
While comparisons conducted using a single stellar evolution library are likely to 
result in robust differences/similarities, it is the absolute time frame that 
will depend on the choice of stellar evolution library.
A goal of the ISLAndS project is to 
conduct comparisons between MW and M31 satellite SFHs using multiple stellar 
evolution libraries.

\clearpage

% LocalWords:  thu SagDIG SFH LGS IAC apa etal PSF DOLPHOT DAOPHOT ALLFRAME CMD
% LocalWords:  PSFs photometries SFHs ACS isochrones BaSTI MSTO overplotted Gyr
% LocalWords:  metallicity HB Lyrae RGB HeB LCID SFR SSP metallicities pCMD CEL
% LocalWords:  sCMD oCMD CMDs pCMDs Montecarlo obsersin unrecovered MinnIAC Myr
% LocalWords:  parameterization CDMs subgiant nB wB sB gB fB dSphs Tucana dSph=
%%
%% table 1

\begin{deluxetable*}{lcccccrrrrr}
\tablewidth{0pt}
\tablecaption{Summary of the ISLAndS Sample and Observations\label{tab:galaxies}}
\tabletypesize{\scriptsize}
\tablecolumns{9}
\tablehead{
\colhead{Galaxy}        	&
%\colhead{R.A.}         		&
%\colhead{Decl.}		     	&
\colhead{HST}			&
\colhead{F475W}       		&
\colhead{F814W}         	&
\colhead{$(m-M)_0$}		&
\colhead{E(B-V)}	        &
\colhead{M$_{V}$}               &
\colhead{R$_{1/2}$}               &
\colhead{D$_{M31}$}             &
\colhead{V$_{c,1/2}$}           &
\colhead{V$_{c,1/2}$}
\\
\colhead{}       		&
%\colhead{J2000}         	&
%\colhead{J2000}         	&
\colhead{ID} 			&
\colhead{(sec)}       		&
\colhead{(sec)}	 		&
\colhead{(mag)}			&
\colhead{(mag)}	                &
\colhead{(mag)}                 &
\colhead{(pc)}                  &
\colhead{(kpc)}                 &
\colhead{km s$^{-1}$}           &
\colhead{km s$^{-1}$}
\\
\colhead{(1)}      		&       
\colhead{(2)}        		&       
\colhead{(3)}      		&       
\colhead{(4)}        		&       
\colhead{(5)}        		&       
\colhead{(6)}          		&        
\colhead{(7)}          		&   
\colhead{(8)}                   &
\colhead{(9)}                   &
\colhead{(10)}                  &
\colhead{(11)}                   
}
\startdata
%And I       & 01:42:17.4 & 26:22:00 & 13739 & 19,833 & 15,709 & $24.36\pm0.07$ & 0.047 & $-$11.7 & 672 & 58   \\
%And II      & 02:00:10.1 & 28:49:52 & 13028 & 22,472 & 17,796 & $24.07\pm0.06$ & 0.063 & $-$12.4 &1176 & 184  \\	   
%And III     & 01:46:42.2 & 26:48:05 & 13739 & 28,996 & 22,968 & $24.37\pm0.07$ & 0.050 & $-$10.0 & 479 & 75  \\     
%And XV      & 01:55:20.2 & 27:57:14 & 13739 & 22,443 & 17,773 & $24.00\pm0.20$ & 0.041 & $-$9.4  & 174 & 176 \\	  
%And XVI     & 01:41:07.6 & 27:19:24 & 13028 & 19,833 & 15,709 & $23.60\pm0.20$ & 0.066 & $-$7.5  & 136 & 279 \\	  
%And XVIII   & 07:36:10.3 & 09:59:11 & 13739 & 26,360 & 20,880 & $24.10\pm0.20$ & 0.080 & $-$8.5  & 270 & 370 \\	 
And I       & 13739 & 19,833 & 15,709 & 24.47 & 0.047 & $-$11.4 & 815 & 68  & 16.1 $\pm$ 4.4       & 18 $\pm$ 4   \\
And II      & 13028 & 22,472 & 17,796 & 24.12 & 0.063 & $-$11.7 & 965 & 195 & 12.3 $\pm$ 2.6      & \nodata      \\   
And III     & 13739 & 28,996 & 22,968 & 24.36 & 0.050 & $-$9.6  & 405 & 86  & 14.7 $\pm$ 3.7      & 16 $\pm$ 2   \\     
And XV      & 13739 & 22,443 & 17,773 & 24.66 & 0.041 & $-$8.7  & 314 & 108 & 6.3 $^{+3.4}_{-3.3}$ & 7 $\pm$ 3 \\  
And XVI     & 13028 & 19,833 & 15,709 & 23.60 & 0.066 & $-$7.5  & 130 & 319 & 8.8 $^{+3.2}_{-2.7}$ & 7 $\pm$ 6  \\  
And XXVIII  & 13739 & 26,360 & 20,880 & 24.35 & 0.080 & $-$8.5  & 270 & 368 & 10.4 $^{+7.7}_{-5.8}$ & 8 $\pm$ 3   
\enddata
\tablecomments{\scriptsize{Column 1$-$Galaxy name. 
%Columns 2 and 3$-$Coordinates of galaxy in J2000. 
Column 2$-$HST observing program. 
Columns 3 and 4$-$Integration time in the F475W and F814W filters with the ACS instrument. 
Column 5$-$Distances derived in this paper \citep[on the TRGB scale of][see text]{rizzi2007}. 
Column 6$-$Galactic absorption from the dust maps of \citet{Schlegel1998} with the recalibration from \citet{Schlafly2011}. 
Column 7$-$Absolute V luminosity calculated from distances derived in this paper and apparent magnitudes from
Martin et al.\ (in prep.) and \citet{mcc2012} (And~XXVIII).
Column 8, 9$-$Half-light radius and distance from M31 from Martin et al.\ (in prep.) and \citet{mcc2012} (And~XXVIII), 
and And~XV corrected to our distance.
Column 10, 11$-$ Circular velocity measured at the half light radius following \citet{walker2009} from 
\citet{collins2014} and \citet{tollerud2014}.
}}
\end{deluxetable*}

%% table 2

\begin{deluxetable*}{lcccccccr}
\tablewidth{0pt}
\tablecaption{ISLAndS Sample Kinematic and Abundance Observations from the Literature\label{tab:kinabs}}
\tabletypesize{\scriptsize}
\tablecolumns{9}
\tablehead{
\colhead{Galaxy}                 &
\colhead{$\sigma_{rv}$}          &
\colhead{N stars}                &
\colhead{$<$[Fe/H]$>$}           &
\colhead{$\sigma_{[Fe/H]}$}      &
\colhead{N stars}                &
\colhead{$<$[$\alpha$/Fe]$>$}    &
\colhead{N stars}                &
\colhead{Reference}               
\\
\colhead{}                 &
\colhead{km s${-1}$}       &
\colhead{}                 &
\colhead{}                 &
\colhead{}                 &
\colhead{}                 &
\colhead{}                 &
\colhead{}                 &
\colhead{}                  
\\
\colhead{(1)}                   &
\colhead{(2)}                   &
\colhead{(3)}                   &
\colhead{(4)}                   &
\colhead{(5)}                   &
\colhead{(6)}                   &
\colhead{(7)}                   &
\colhead{(8)}                   &
\colhead{(9)}                   
}
\startdata
And I       & 10.6 $\pm$ 1.1 & 80      & -1.45 $\pm$ 0.04 & 0.37    &  80      & \nodata         & \nodata & \citet{kalirai2010}   \\
And I       & 10.2 $\pm$ 1.9 & 51      & \nodata          & \nodata &  \nodata & \nodata         & \nodata & \citet{tollerud2012}   \\
And I       &  8.2 $\pm$ 1.7 & 49      & \nodata          & \nodata &  \nodata & \nodata         & \nodata & \citet{tollerud2012}  \\
And I       & \nodata        & \nodata & -1.11 $\pm$ 0.12 & \nodata &  31      & 0.28 $\pm$ 0.16 & 7       & \citet{vargas2014}   \\
\\
And II      & 7.3 $\pm$ 0.8  & 95      & -1.64 $\pm$ 0.04 & 0.34    &  95      & \nodata         & \nodata & \citet{kalirai2010}   \\
And II      & 7.8 $\pm$ 1.1  & 531     & -1.39 $\pm$ 0.03 & 0.72 $\pm$ 0.03    &  477  & \nodata & \nodata & \citet{ho2012}   \\
And II      & \nodata        & \nodata & -1.37 $\pm$ 0.12 & \nodata &  248     & 0.03 $\pm$ 0.09 & 56      & \citet{vargas2014}   \\
And II      & \nodata        & \nodata & -1.25 $\pm$ 0.05 & 0.49 $\pm$ 0.04    &  300  & \nodata & \nodata & \citet{ho2015}   \\
\\
And III     & 4.7 $\pm$ 1.8  & 43      & -1.78 $\pm$ 0.04 & 0.27    &  43      & \nodata         & \nodata & \citet{kalirai2010}  \\
And III     & 9.3 $\pm$ 1.4  & 62      & \nodata          & \nodata &  \nodata & \nodata         & \nodata & \citet{tollerud2012}  \\
And III     & \nodata        & \nodata & -1.81 $\pm$ 0.12 & \nodata &  35      & 0.33 $\pm$ 0.21 & 8       & \citet{vargas2014}   \\
\\
And XV      & 4.0 $\pm$ 1.4  & 29      & \nodata          & \nodata &  \nodata & \nodata         & \nodata & \citet{tollerud2012}  \\
\\
And XVI     & 3.8 $\pm$ 2.9  & 7       & \nodata          & \nodata &  \nodata & \nodata         & \nodata & \citet{tollerud2012}   \\
And XVI     & 5.8 $^{+1.1}_{-0.9}$ & 20   & -2.0 $\pm$ 0.1  & \nodata &  12    & \nodata         & \nodata & \citet{collins2015}   \\
\\
And XXVIII  & 4.9 $\pm$ 1.6  & 18      & $\sim$-2.0        & \nodata &  \nodata & \nodata         & \nodata & \citet{tollerud2013}  \\
And XXVIII  & 6.6 $^{+2.9}_{-2.1}$ & 17   & -2.1 $\pm$ 0.3  & \nodata &  17     & \nodata         & \nodata & \citet{collins2013}   \\
And XXVIII  &  \nodata       & \nodata & -1.84 $\pm$ 0.15  & 0.65 $\pm$ 0.15   &  13 & \nodata    & \nodata & \citet{slater2015} 
\enddata
\tablecomments{\scriptsize{Column 1$-$Galaxy name. Column 2$-$Line of sight velocity dispersion.
Column 3$-$Number of stars used to calculate the velocity dispersion.
Column 4$-$Average metallicity.
Column 5$-$Dispersion in metallicity.
Column 6$-$Number of stars used to calculate metallicity.
Column 7$-$Average alpha element abundance relative to metallicity
Column 8$-$Number of stars used to calculate average alpha element abundance.
Column 9$-$Reference for literature source.
}}
\end{deluxetable*}

%% table 3

\begin{deluxetable*}{lcccccc}
\tablewidth{0pt}
\tablecaption{Summary of the ISLAndS Sample Distances and Intrinsic Properties\label{tab:distances}}
\tabletypesize{\scriptsize}
\tablecolumns{6}
\tablehead{
\colhead{Galaxy}                &
\colhead{TRGB}                   &
\colhead{A$_I$}                 &
\colhead{$(m-M)_0$}             &
\colhead{m$_{V,0}$}                &
\colhead{M$_{V}$}               &
\colhead{M$_{tot,1/2}$} 
\\
\colhead{}                      &
\colhead{F814W}                    &
\colhead{(mag)}                 &
\colhead{(mag)}                 &
\colhead{(mag)}                 &
\colhead{(mag)}                 &
\colhead{(M$_\odot$)}               
\\
\colhead{(1)}                   &
\colhead{(2)}                   &
\colhead{(3)}                   &
\colhead{(4)}                   &
\colhead{(5)}                   &
\colhead{(6)}                   &
\colhead{(7)}            
}
\startdata
And I       & 20.50 $\pm$ 0.01 & 0.080 & 24.47 & 13.1 $\pm$ 0.1 & $-$11.4  & 3.2 $\times$ 10$^7$ \\
And II      & 20.16 $\pm$ 0.01 & 0.092 & 24.12 & 12.4 $\pm$ 0.1 & $-$11.7  & 3.4 $\times$ 10$^7$  \\
And III     & 20.39 $\pm$ 0.02 & 0.084 & 24.36 & 14.8 $\pm$ 0.1 & $-$9.6   & 2.0 $\times$ 10$^7$  \\
And XV      & 20.68 $\pm$ 0.08 & 0.070 & 24.66 & 16.0 $\pm$ 0.1 & $-$8.7   & 2.9 $\times$ 10$^6$  \\
And XVI     & \nodata          & \nodata & 23.60 & 16.1 $\pm$ 0.1 & $-$7.5 & 2.5 $\times$ 10$^6$   \\
And XXVIII  & \nodata          & \nodata & 24.35 & 15.9 $\pm$ 0.5 & $-$8.5 & 6.8 $\times$ 10$^7$ 
\enddata
\tablecomments{\scriptsize{Column 1$-$Galaxy name.
Column 2$-$TRGB measured from HST observations. 
Column 3$-$I-band Galactic absorption from \citet{Schlafly2011}.
Column 4$-$Distance modulus from columns 1 and 2. For And~XV and And~XVIII, distances were derived 
from the best solutions from MATCH. 
Column 5$-$Extinction corrected V-band apparent magnitude from Martin et al.\ (in prep.).
Column 6$-$Absolute V-band magnitude.
Column 7$-$Total mass within the half-light radius (from Table~1) and the stellar velocity dispersions 
(latest values from Table~2) and using the mass estimator from \citet{walker2009}.
}}
\end{deluxetable*}

\end{document}